\documentclass[%
 reprint,
 superscriptaddress,
 floatfix,
 amsmath, amssymb,
 aps,
 prx,
]{revtex4-2}
\usepackage[utf8]{inputenc}
\usepackage[english]{babel}
\usepackage[T1]{fontenc}
\usepackage{amsmath}

\usepackage[]{hyperref}

\usepackage{multirow}
\usepackage{booktabs}

\usepackage{amsmath,amssymb,amsthm,bm,amsfonts,mathrsfs,bbm}

\DeclareMathOperator*{\argmin}{arg\,min}

\usepackage{dsfont}

\usepackage{graphicx} 
\usepackage{float}

\makeatletter
\let\newfloat\newfloat@ltx
\makeatother
\usepackage{algorithm}

\usepackage[noend]{algpseudocode}

\usepackage{bbold}
\usepackage{braket}

\usepackage{cleveref}

\usepackage{changes}

\usepackage{xcolor}

\begin{document}

\title{Reducing Entanglement With Physically-Inspired Fermion-To-Qubit Mappings}

\author{Teodor Parella-Dilmé}
\email[E-mail:]{teodor.parella@icfo.eu}
\affiliation{ICFO - Institut de Ciencies Fotoniques, The Barcelona Institute of Science and Technology, Av. Carl Friedrich Gauss 3, 08860 Castelldefels (Barcelona), Spain}

\author{Korbinian Kottmann}
\affiliation{ICFO - Institut de Ciencies Fotoniques, The Barcelona Institute of Science and Technology, Av. Carl Friedrich Gauss 3, 08860 Castelldefels (Barcelona), Spain}

\author{Leonardo Zambrano}
\affiliation{ICFO - Institut de Ciencies Fotoniques, The Barcelona Institute of Science and Technology, Av. Carl Friedrich Gauss 3, 08860 Castelldefels (Barcelona), Spain}

\author{Luke Mortimer}
\affiliation{ICFO - Institut de Ciencies Fotoniques, The Barcelona Institute of Science and Technology, Av. Carl Friedrich Gauss 3, 08860 Castelldefels (Barcelona), Spain}

\author{Jakob~S.~Kottmann}
\email[E-mail:]{jakob.kottmann@uni-a.de}
\affiliation{{Institute for Computer Science, University of Augsburg, Germany }}
\affiliation{{Center for Advanced Analytics and Predictive Sciences, University of Augsburg, Germany }}

\author{Antonio Acín}
\affiliation{ICFO - Institut de Ciencies Fotoniques, The Barcelona Institute of Science and Technology, Av. Carl Friedrich Gauss 3, 08860 Castelldefels (Barcelona), Spain}
\affiliation{ICREA-Institucio Catalana de Recerca i Estudis Avancats, Lluis Companys 23, 08010 Barcelona, Spain}

\date{\today}

\begin{abstract}
In ab-initio electronic structure simulations, fermion-to-qubit mappings represent the initial encoding step of the fermionic problem into qubits. This work introduces a physically-inspired method for constructing mappings that significantly simplify entanglement requirements when simulating states of interest. The presence of electronic excitations drives the construction of our mappings, reducing correlations for target states in the qubit space. To benchmark our method, we simulate ground states of small molecules and observe an enhanced performance when compared to classical and quantum variational approaches from prior research employing conventional mappings. In particular, on the quantum side, our mappings require a reduced number of entangling layers to achieve accuracy for $LiH$, $H_2$, $(H_2)_2$, the $H_4^{\neq}$ stretching and benzene's $\pi$ system using the RY hardware efficient ansatz. In addition, our mappings also provide an enhanced ground state simulation performance
in the density matrix renormalization group algorithm for the $N_2$ molecule.
\end{abstract}

\maketitle
\section{Introduction}
Electronic structure ab-initio quantum chemistry simulations play a pivotal role in various fields, including drug development \cite{doi:10.1021/acs.jctc.2c00574, 8585034}, computational catalysis \cite{von_Burg_2021}, and material science \cite{Lordi_Nichol_2021}. Given the significance of this domain, numerous techniques have evolved over the years, enabling simulations of increasingly complex systems. In the context of our research, it is worth mentioning three specific methodologies: classical simulations, variational quantum algorithms executed on current and near-term noisy quantum devices, and fault-tolerant quantum computing. Classical simulations, in form of the density matrix renormalization group algorithm (DMRG) \cite{white1992density, schollwock2005density,chan2011density, Baiardi_2020} and Monte-Carlo methods \cite{hammond1994monte}, stand as state of the art methodologies and are readily available. Variational quantum algorithms are executed on proof-of-concept scales on contemporary noisy quantum computing devices, and hold the potential to achieving simulations of useful system sizes in the near future. Finally, quantum algorithms on fault tolerant quantum computers are projected to yield useful applications in a farther future where such computers are available, and the gate counts and runtimes of these algorithms reduced to practical numbers. 
\\
\\
Central to all approaches is the transformation of the physical problem into units of information. This involves the selection of an appropriate basis consisting in a set of spin-orbitals (SOs), on top of a fermion-to-qubit mapping. In this work, we focus on the choice of fermion-to-qubit mappings, which leads to potential advantages in all three formerly mentioned avenues. 
\\
\\
Fermion-to-qubit mappings represent the initial encoding step of fermionic problems into qubits, faithfully replicating the algebra of fermionic operators on a qubit system \cite{BRAVYI2002210}. Given the non-local nature of the fermionic state space, these mappings introduce a non-local structure into qubit systems, which are typically characterized by a limited connectivity. It is crucial to emphasize that these mappings are not unique, as there are numerous possible solutions to the encoding problem. Consequently, depending on the qubit topology and the applied technique, some mappings may prove more advantageous than others for simulation tasks.  In this context, the reduction of Pauli weight, mapped operators locality, entangling gate overhead and number of qubits has played a leading role in the design of novel mappings, all while considering the practical constraints imposed by experimental hardware \cite{Vlasov_2022,jiang2020optimal, steudtner2018fermion, Verstraete_2005, PhysRevB.104.035118, chien2019analysis, chien2022optimizing, miller2022bonsai}. Here, we follow a different path and choose entanglement as the relevant figure of merit: our goal is to design fermion-to-qubit mappings that reduce the entanglement in the target
qubit state.
\\
\\
Given an initial fermionic Hamiltonian, its energy spectrum remains invariant under any valid fermion-to-qubit transformation. However, the eigenstates of the resulting qubit Hamiltonian, and consequently, the target state $|\psi\rangle$ to be simulated, differs across the mappings. In particular, it can exhibit substantial variations on its entanglement properties in qubit representation, which is a critical factor influencing the quality of the simulations, both in the classical and quantum case. For instance, it is not clear if the Pauli weight has any
impact on DMRG simulations, while they are known to be affected by the entanglement structure of the target state. In general, identifying mappings yielding to states with reduced entanglement requirements could substantially enhance simulation performance.
\\
\\
In this work, a physically-inspired method is proposed to construct tailored entanglement-aware fermion-to-qubit mappings with reduced entanglement requirements on the state being simulated. For small molecular systems, we prove enhanced ground-state preparation performance respect to paradigmatic mappings. In the quantum computation framework, we use the variational quantum eigensolver (VQE) with an RY hardware efficient ansatz (HEA) to demonstrate improved accuracy for $LiH$, $H_2$, $(H_2)_2$, the $H_4^{\neq}$ stretching and benzene's $\pi$ system.
Additionaly, we extend the applicability of our method to the tensor networks framework, where we simulate $N_2$ using DMRG on a matrix product state (MPS) representation.


\section{Fermion-to-qubit mappings}
\label{section2}

The problem of designing beneficial qubit representations of electronic systems is rich and can
be tackled from different angles. As mentioned, in this work we are focusing on the fermion-to-qubit mapping alone. Other works employ different strategies to reach this goal, such as optimizing the spatial orbitals~\cite{Mitarai2018, sokolov2020quantum, kottmann2023molecular} or regularizing the electronic Hamiltonian~\cite{motta2020quantum, mcardle2020improving, schleich2022improving, kumar2022quantum, sokolov2023orders, dobrautz2023ab, volkmann2023qubit}, among others~\cite{PhysRevB.92.075132,Robin_2021,Robin_2023, Hengstenberg_2023}. Note that these are separate strategies that can be combined with optimized fermion-to-qubit mappings.\\

The proposed mappings can be related to recent works aiming to find optimal Clifford circuits for entanglement reduction ~\cite{mishmash2023hierarchical, schleich2023partitioning, Shang_Chen_Yuan_Lu_Pan_2023}. These works can be interpreted as finding a beneficial mapping through a transformation included in the quantum circuit, while in this work we aim to find a suitable mapping directly. Our results therefore apply to both classical
and quantum approaches.

\subsection{Fock Space to Qubits}
The antisymmetrized Fock space $\mathcal{F}_{-}$ encompasses all conceivable states from a many-body fermionic system. In electronic structure problems, solving for the spectrum of an operator $\hat{O} \in Lin[\mathcal{F}_{-}]$ holds significant importance, a particular case being the electronic Hamiltonian. The fermionic algebra consists in a set of creation $\{a_i^{\dagger}\}_{i=1}^n$ and annihilation $\{a_i\}_{i=1}^n$ operators in $Lin[\mathcal{F}_{-}]$, satisfying Canonical Anticommutation Relations (CAR):
\begin{equation}
    \{a_i^\dagger,a_j\}=\delta_{ij}\mathds{1} \hspace{.5cm},\hspace{.5cm} \{a_i,a_j\}=\{a_i^\dagger,a_j^\dagger\}=0 \hspace{.25cm}.
    \label{CAR}
\end{equation}
When applied to the fermionic vacuum $|vac\rangle$, such ladder operators define the orthonormal Fock basis spanning $\mathcal{F}_-$, in the so-called occupation number vectors notation
\begin{equation}
    |f_1,f_2,...,f_n \rangle \equiv 
    (a_1^\dagger)^{f_1}(a_2^\dagger)^{f_2}\hspace{.1cm}...\hspace{.1cm}(a_n^\dagger)^{f_n}|vac\rangle \hspace{.25cm}.
    \label{ONV}
\end{equation}
Here, $f_i \in\{0,1\}$ is called the fermionic occupation of mode $i$, $(a_i^\dagger)^0 \equiv \mathds{1}$ and $n$ is the total number of fermionic modes available. Note that the ladder operators are applied in descending order, as a phase factor could apply otherwise. Equivalently, ladder operators can be decomposed as
\begin{equation}
    a_j=\frac{1}{2}(\gamma_{2j}+i\gamma_{2j-1})\hspace{.5cm},\hspace{.5cm}
    a_j^\dagger=\frac{1}{2}(\gamma_{2j}-i\gamma_{2j-1}) \hspace{.25cm},
    \label{MajoranaDecomposition}
\end{equation}
into a self-adjoint part ($\gamma_{2j}$) and an anti-self-adjoint part ($i\gamma_{2j-1}$). They are uniquely identified by a Majorana basis $\{\gamma_j\}_{j=1}^{2d}$, satisfying both being self-adjoint ($\gamma_j=\gamma_j^\dagger$) and the associated CAR for Majoranas
\begin{equation}
    \{\gamma_j,\gamma_k\}=2\delta_{jk}\mathds{1} \hspace{.25cm}.
    \label{majoranaCAR}
\end{equation}
Operators in $\it{Lin}[\mathcal{F}_{-}]$ can be expressed as linear combinations of products of a set of ladder operators. In particular, the electronic Hamiltonian is expressed as a linear combination of one-body and two-body operators as
\begin{equation}
    \hat{H}_{e}=\sum_{i,j} h_{ij}  \hat{a}_i^{\dagger}\hat{a}_j+ \sum_{i,j,k,l}h_{ijkl} \hat{a}_i^{\dagger}\hat{a}_j^{\dagger}\hat{a}_l \hat{a}_k \hspace{.25cm},
    \label{2ElectronicHamiltonian}
\end{equation}
with $h_{ij}$,  $h_{ijkl}$ as the one and two-body terms respectively, available from classical computation methods. 
\\
\\
The abstract Fock space spanned by $n$ fermionic modes is isomorphic to the Hilbert space spanned by $n$ qubits $\mathcal{F}_- \simeq (\mathbb{C}^{2})^{\otimes n}$, with dimension $2^n$. Hence, it is always
possible to transform the Fock space to a qubit space through a so-called fermion-to-qubit mapping:
\begin{equation}
\Lambda\hspace{.1cm}:\hspace{.5cm}\mathcal{F}_- \hspace{.1cm}\xrightarrow{}\hspace{.1cm} (\mathbb{C}^{2})^{\otimes n}\hspace{.25cm}.
     \label{F2Qmap}
\end{equation}
These mappings also transform ladder operators into operators in $Lin[\mathcal{F}_{-}]$ satisfying the CAR. After application of any of such mappings, any electronic Hamiltonian $ \hat{H}_{e}$ is fully reconstructed in the qubit space.


\subsection{Ternary Tree Encodings}
\begin{figure}[t]                     
\centering
\includegraphics[width=1\columnwidth, angle=0]{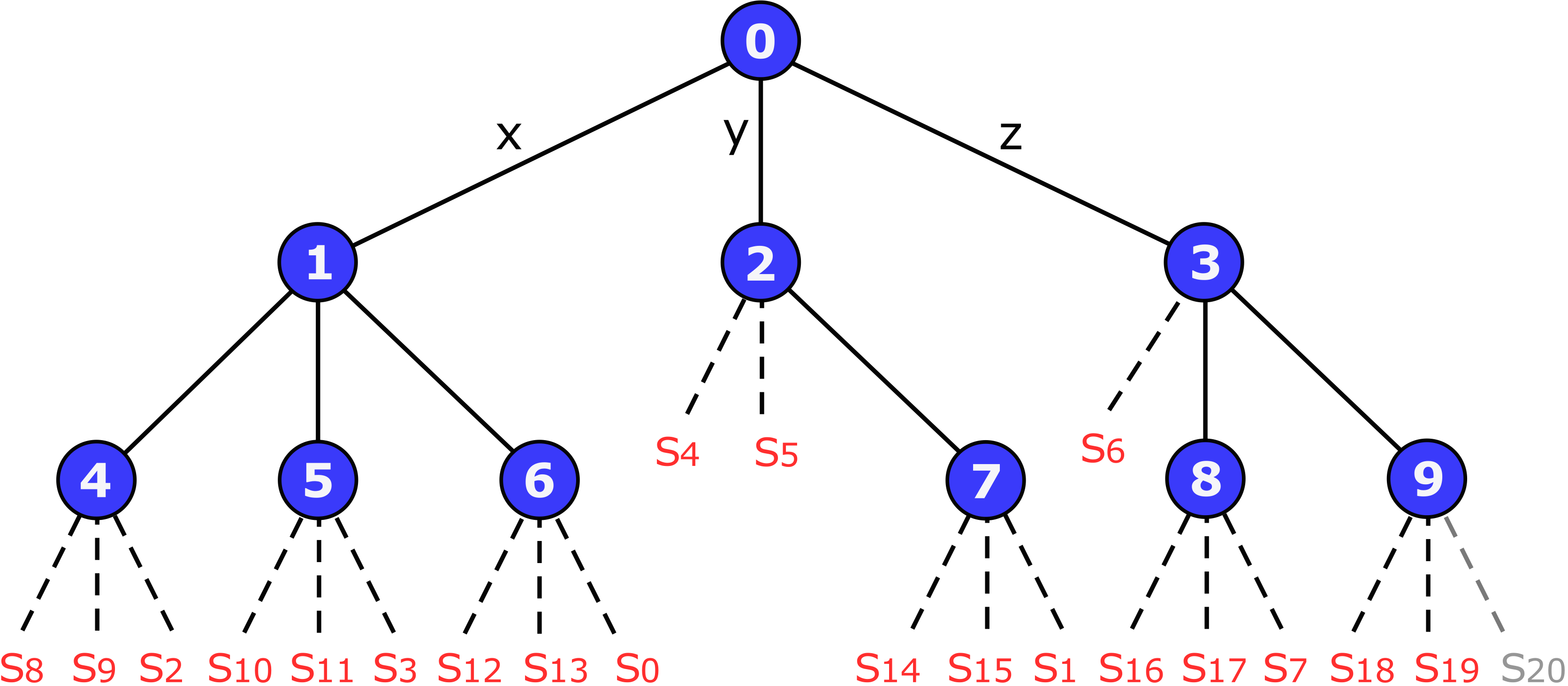}
\caption{Ternary Tree example, corresponding to a fermion-to-qubit mapping of 10 fermionic modes. Each node holds a fermionic mode, an ascending parent node (except from the root node at the top), and three descending labeled  branches $x$,$y$,$z$ (from left to right respectively) indicating the child nodes. The dashed lines correspond to legs of the tree, each retrieving a single Majorana string $S_i$ in red. The leg associated to $z$-paths form the root node is marked in grey, and its Majorana string is unpaired.}
\label{fig:ExampleTT}
\end{figure}

The relation between ternary trees (TT) and mappings was introduced in \cite{Vlasov_2022}. In \cite{jiang2020optimal}, a TT mapping was proven optimal to minimise the Pauli weight of the derived operators in $Lin[\mathcal{F}_{-}]$. This ignited the research in \cite{miller2022bonsai}, where the flexibility of generating TT mappings is implemented favourably to experimental hardware. Moreover, the last reference includes a thorough discussion on TT mappings, a detailed pairing scheme preserving fermionic vacuum, and analysis of occupation localisation. Overall, the TT formalism provides a large framework of fermion-to-qubit mappings based on Pauli strings, including the representation of traditional mappings such as Jordan-Wigner (JW) \cite{jordan1993paulische}, Parity and Fenwick tree based mappings like Bravyi-Kitaev \cite{BRAVYI2002210}. Such TT formalism enables the retrieval of a Majorana basis set in the qubit representation from a TT, satisfying Eq.(\ref{majoranaCAR}).
\\
\\
We briefly introduce a TT as a graph $\mathcal{T}=(V,E)$ consisting of a set of $m$ vertices $V=\{v_i\}_{i=1}^{m}$ and $(m-1)$ connecting edges $E\subseteq\{(x,y)\in V^2|x\neq y\}$ [Fig.\ref{fig:ExampleTT}]. In the upper part of the TT there is the root node, which ramifies the graph iteratively up to 3 labeled descendants per node. For two connected nodes $v_i$, $v_j$, a kinship relationship is set between them. With respect to $v_i$, the node $v_j$ can take the relationship of parent ($v_i^p$), $x$-child ($v_i^x$), $y$-child ($v_i^y$) or $z$-child ($v_i^z$). The same applies for $v_i$. Except for the root node (which has no parent), every node has one and only one associated parent, and up to three labeled children. If a node is missing some $x$, $y$ or $z$ child, then for each missing child a leg from the ternary tree is defined. Formally, a leg refers to an edge that is connected to a single graph node, with no connection at its other end. In total, there are $2m+1$ resulting legs on the ternary tree. Each leg $l$ has an associated Pauli string $S_l$, which can be retrieved by tracing the ascending path back to the root node. Initially, the node $v_i$ on top of the current leg $x/y/z$ denotes an $X/Y/Z$ Pauli operator to be applied to the $i$th qubit. Subsequently, as we progress to the parent node $v_j = v_i^p$, the relationship between $v_i$ and $v_j$ delineates $v_j^x/v_j^y/v_j^z$, signifying an $X/Y/Z$ Pauli operator to be applied to the $j$th qubit. By redefining $v_i \equiv v_i^p$, the process is repeated iteratively, where each iteration retrieves the operators to be applied to each parent node. In the final step, $v_i^p$ becomes the root node of the ternary tree, having no associated parents, and terminating the process. The associated Pauli string $S_l$ for leg $l$ results from tensoring the Pauli operators of the aforementioned progression. The resulting set of Pauli strings from the TT are in fact defined as Majorana strings ($S_l$=$S_l^\dagger$). These replicate the fermionic CAR for Majoranas from Eq.(\ref{majoranaCAR}) in qubit space. Ultimately, the Majorana strings may be paired in order to preserve the fermionic vacuum through the pairing algorithm [Alg.\ref{AlgorithmPairing}, Appendix \ref{appendixA}]. 
\\
\\
For further details about the TT formalism, we refer the reader to the detailed explanations in \cite{Vlasov_2022, miller2022bonsai}. In this
work, the TT formalism has been implemented using the \textsc{tequila} \cite{kottmann2021tequila} Python library.

\subsection{Fermionic Permutation and Qubit Permutation}

\begin{figure} [t]                    
\centering
\includegraphics[width=1\columnwidth, angle=0]{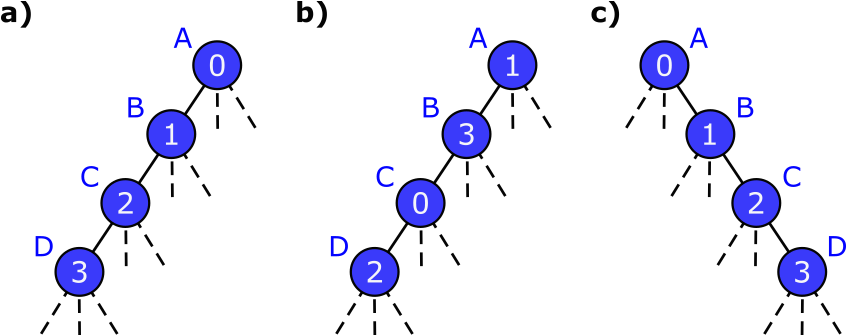}
\caption{Ternary trees of fermion-to-qubit encodings with 4 fermionic modes (0,1,2,3) into 4 qubits (A,B,C,D). \textbf{a)} Parity encoding of fermions in standard order (0,1,2,3). \textbf{b)} Parity encoding of permuted fermions in order (1,3,0,2). \textbf{c)} JW encoding in standard order.}
\label{fig:mapping_example}
\end{figure}

In this section, our aim is to discuss a fundamental yet often misunderstood distinction between fermionic permutation and qubit permutation. The former refers to the relative permutation of fermionic modes encoded within the TT mapping, directly affecting the transformation and further simulation requirements \cite{Chiew2023discoveringoptimal}. The latter involves a reordering of qubits, leaving the transformation intact but bearing significant implications when conducting simulations with algorithms or devices with restricted topologies. To ensure clarity, we elaborate on the distinction between both permutations through explicit and illustrative examples. 
\\
\\
We draw special attention to the resulting effects of encoding parity locally. In the literature, parity mappings are often referred to as a unique parity mapping. However, the specific order in which the fermionic occupation is encoded results in non-equivalent mappings.
\\
\\
Encoding locally in parity involves encoding in either the $x$-branch or the $y$-branch of a ternary tree. To simplify our discussion, we will focus on the former case and consider an $x$-branch mapping of 4 fermionic modes labeled 0,1,2,3, onto four qubits labeled A,B,C,D in standard order [Fig.\ref{fig:mapping_example}a]. We will explore another $x$-branch parity mapping where modes are encoded in the order 1,3,0,2 [Fig.\ref{fig:mapping_example}b]. Lastly, we present the JW mapping in the standard order for the purpose of comparison [Fig.\ref{fig:mapping_example}c].
\\
\\
The fermionic modes held by each node from the tree do have relevant implications on the mapping, and do in fact lead to different fermion-to-qubit mappings. We say that the two parity mappings presented differ due to a permutation in the fermionic space. Considering the first mapping [Fig.\ref{fig:mapping_example}a], qubit D holds the occupation of fermionic mode 3, qubit C the total occupation sum modulo 2 of modes 2 and 3, qubit B the total occupation sum modulo 2 of modes 1,2 and 3, and qubit A the total occupation sum modulo 2 of all 4 modes. However, in the case of mapping [Fig.\ref{fig:mapping_example}b], qubit D holds the occupation of fermionic mode 2, qubit C the occupation modulo 2 of modes 0 and 2, qubit B the occupation modulo 2 of modes 3, 0 and 2, and qubit A the occupation modulo 2 of all 4 modes. Both mappings have qubits that encode different occupation information from any other qubit in the other mapping. For instance, in the first mapping there is a qubit holding occupation information from mode 3, or a qubit holding occupation modulo 2 of modes 2 and 3. However, in the second parity mapping there are no such qubits, instead they are different. They are essentially non-equivalent fermion-to-qubit mappings, leading to alternative representations of the Fock space in the qubit space. 
\\

On top of this fermionic permutation that changes the mapping, one can define the physical qubit permutation, related to the relative order of the qubits A,B,C,D. Such physical permutation matters when the order of entangled subparties is relevant within an optimization method. For instance, physical permutation is important within a VQE optimization in a device with restricted qubit connectivity, or in a specific topological decomposition of a tensor (such as in a linear MPS). We remark that the physical permutation does not modify the representation of a state, but it has implications on optimization methods with finite resources.


\section{Simplifying Entanglement by Encoding Ocupation Non-Locally}
\label{section3}

The study of many-body electronic ground states involves dealing with complex correlations. Typically, a Hartree-Fock (HF) self-consistent field approach is used in first instance to identify an optimized separable state. However, to account for correlations, it is essential to consider electronic excitations, which entail promoting electrons from occupied SOs to virtual ones. Currently, widely-used methods like configuration interaction \cite{langhoff1974configuration} and coupled cluster \cite{bartlett2007coupled} aim to represent these excitations through parameterizations. The principle remains similar in its unitary variant (UCC) (see section ~\ref{ref:sec-Reducing Entanglement on Small Molecular Systems})
~\cite{anand2022quantum}.
\\
\\
In this section, we propose an approach to streamline the treatment of correlations during the fermion-to-qubit encoding process. To illustrate this idea, we initially examine idealized fermionic scenarios involving pure states.

\subsection{Double electronic excitation}
We first consider a scenario involving four SOs related to a generalized double excitation, which corresponds to the simultaneous promotion of two electrons from two occupied fermionic modes to two unoccupied ones. An example could be a simultaneous promotion of two electrons occupying the two SOs from the same MO to the two SOs from a virtual MO.  In the context of the JW encoding [Fig.\ref{fig:mapping_example}c], the occupation information of the initially-filled fermionic modes 0 and 1 is locally encoded in qubits A and B respectively. Similarly, the occupation of the two virtual fermionic modes 2 and 3 is locally encoded in qubits C and D. In this context, the state resulting from the double excitation is represented as follows:
\begin{equation}
    |\psi\rangle^{JW}_{ABCD} = \alpha |1100\rangle^{JW}_{ABCD} + \beta |0011\rangle^{JW}_{ABCD}\hspace{.25cm}.
    \label{DoubleExcitation_JW}
\end{equation}
An alternative approach may be explored by encoding locally in parity. We call $P$ the parity mapping from [Fig.\ref{fig:mapping_example}a], where qubits A/B/C/D hold respectively the occupation modulo 2 of the last 4/3/2/1 fermionic modes. The basis elements from the JW transformation
\begin{equation}
    \{|1100\rangle^{JW}_{ABCD},|0011\rangle^{JW}_{ABCD}\}\hspace{.25cm},
\end{equation}
are, under the encoding $P$, represented as
\begin{equation}
    \{|0100\rangle^{P}_{ABCD},|0001\rangle^{P}_{ABCD}\}\hspace{.25cm}.
    \label{new_parity_basis}
\end{equation}
That is, the parity map goes from left to right, encoding the occupation modulo 2 of the qubits in the right in JW encoding. Ultimately, the doubly-excited state in Eq.(\ref{DoubleExcitation_JW}) transformed from JW to the $P$ encoding is
\begin{equation}
    \begin{split}
        |\psi\rangle^{P}_{ABCD} &=\alpha |0100\rangle^{P}_{ABCD} + \beta |0001\rangle^{P}_{ABCD}= \\
           &=|00\rangle^{P}_{AC}  \otimes \bigg(\alpha |10\rangle^{P}_{BD} + \beta |01\rangle^{P}_{BD}\bigg)\hspace{.25cm}.
           \label{DoubleExcitation_P}
    \end{split}
\end{equation}
The entangled double-excitation subspace in JW representation involving four qubits from Eq.(\ref{DoubleExcitation_JW}) is effectively reduced to two entangled qubits in Eq.(\ref{DoubleExcitation_P}) using the $P$ encoding. 

\subsection{Single electronic excitation}
Although double excitations represent the major source of correlations in electronic structure problems, it is worth investigating the single electronic excitation case, involving the promotion of a single electron. In fact, the resulting state is symmetric in terms of spin, and single excitations may be computed between two molecular orbitals (MOs), involving a total of 4 SOs. 
\\
\\
Let's explore this concept further by introducing an idealized scenario that revolves around the excitation subspace in JW representation [Fig.\ref{fig:mapping_example}c]. To better understand this, let's take the state $|1100\rangle_{ABCD}$ and possible single-excitations acting on it symmetrically in terms of spin. The resulting subspace is then spanned by $|1100\rangle_{ABCD}$, $|0110\rangle_{ABCD}$ and $|1001\rangle_{ABCD}$. This implies that the state can be represented as follows:
\begin{equation}
    |\psi\rangle^{JW}_{ABCD} = \alpha |1100\rangle^{JW}_{ABCD} + \beta |0110\rangle^{JW}_{ABCD}+ \beta |1001\rangle^{JW}_{ABCD}
    \label{SingleExcitation_JW}
\end{equation}
Upon closer examination, the single excitation seems to be more intricate than initially expected, involving entanglement among all four qubits. We can consider now the mapping $P'$ from [Fig.\ref{fig:mapping_example}b]. The basis elements
\begin{equation}
    \{|1100\rangle^{JW}_{ABCD},|0110\rangle^{JW}_{ABCD},|1001\rangle^{JW}_{ABCD}\}\hspace{.25cm},
\end{equation}
are transformed respectively into
\begin{equation}
    \{|0110\rangle^{P'}_{ABCD},|0111\rangle^{P'}_{ABCD},|0010\rangle^{P'}_{ABCD}\}\hspace{.25cm}.
\end{equation}
The singly-excited state from Eq.(\ref{SingleExcitation_JW}) in JW representation, is transformed in the $P'$ representation to
\begin{equation}
    |\psi\rangle^{P'}_{ABCD} = |01\rangle^{P'}_{AC} \otimes \bigg( \alpha|10\rangle^{P'}_{BD} +\beta|11\rangle^{P'}_{BD}+\beta|00\rangle^{P'}_{BD} \bigg) \hspace{.25cm}.
    \label{SingleExcitation_P}
\end{equation}
Most importantly, the requirement of entangling 4 qubits in JW for Eq.(\ref{SingleExcitation_JW}), is again reduced to 2 qubits in the $P'$ encoding for Eq.(\ref{SingleExcitation_P}).
\\

Note that we have used a CI-like scenario in the previous example, where a potential single electron excitation acts linear on the initial state $\ket{1100}^{JW}_{ABCD}$. In a typical quantum computing scenario, the single excitations will be introduced via a unitary-coupled cluster type operator (see next section) that will also produce the state $\ket{0011}^{JW}_{ABCD}$ where both single electrons are excited at the same time. The corresponding amplitude will be on the order of $\alpha \beta$, so that the effect of the now imperfect compression under the parity mapping will become negligible for small amplitudes. As single electron excitations can be identified with orbital rotations (see Eq.(12) in ~\cite{kottmann2023compact}) this motivates future combinations with orbital optimization techniques.
\\
\\
In complex quantum-chemistry scenarios with multiple SOs and correlations, the reduced density matrices may exhibit a high degree of mixed states. This arises due to the possibility of multiple excitations affecting the same SOs, thereby introducing complexity in the previously idealized cases, even in an potential orbital optimized variant. Although we are constrained in the number of excitations we can simplify using the aforementioned methods, we can still deal with excitations that contribute the most to correlation effects.
\\
\\
In the upcoming sections, our focus will be on identifying relevant excitations and explore how the entanglement map transforms under the fermion-to-qubit mapping chosen. Our objective is to investigate the potential to mitigate correlations by employing a clever mapping strategy in complex scenarios.

\section{Reducing Entanglement on Small Molecular Systems}\label{ref:sec-Reducing Entanglement on Small Molecular Systems}

\begin{figure*}                     
\centering
\includegraphics[width=2\columnwidth, angle=0]{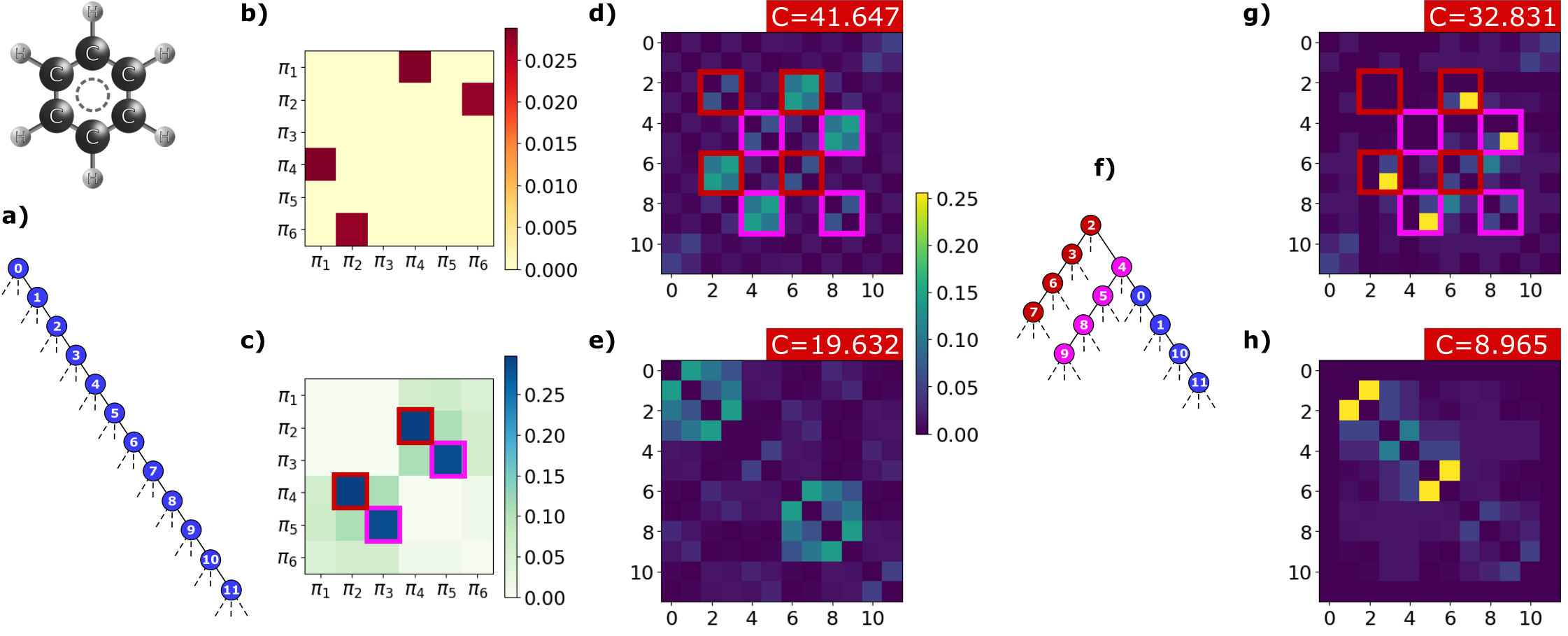}
\caption{Simplification of correlations for the 6 MOs involved in the $\pi$ system of benzene [Fig.\ref{fig:Benzene_Orbitals}, Appendix.\ref{appendixA}]. Red labels indicate the cost function value from Eq.(\ref{CostFunction}) of the corresponding mutual information matrices. Red and pink squares indicate the elements associated to the most relevant double excitations. \textbf{a)} TT corresponding to JW encoding \textbf{b)} Absolute values $|\theta_{ij}^s|$ after UpCCGSD, corresponding to single excitations between MOs $\pi_i$ and $\pi_j$. \textbf{c)} Absolute values $|\theta_{ij}^d|$ after UpCCGSD, corresponding to double excitations between MOs $\pi_i$ and $\pi_j$. \textbf{d)} Mutual information matrix between qubits for JW exact ground state. \textbf{e)} Mutual information matrix reordered minimizing the correlation cost function $C$ (final value displayed). \textbf{f)} Constructed mapping encoding the most relevant double excitation subspaces locally in parity. \textbf{g)} Exact ground state mutual information matrix for the adapted mapping that reduces entanglement on the double excitations between qubits 2,3,6,7 and qubits 4,5,8,9. \textbf{h)} Reordered mutual information matrix for the adapted mapping.}
\label{fig:Benzene_SD}
\end{figure*} 

In order to simplify qubit space entanglement as outlined in the previous section, we first proceed identify the primary sources of correlation based on single and/or double excitations. Then, we will replicate the entanglement streamline procedure by locally encoding parity of the fermionic modes associated to the selected excitation(s).  For achieving these objectives, the algorithmic steps implemented in this work are detailed in [Alg.\ref{TailoredMappings}].
\\
\\
For the sake of clarity, we provide a detailed implementation of [Alg.\ref{TailoredMappings}] along the illustrative example of benzene's $\pi$ system [Fig.\ref{fig:Benzene_Orbitals}, Appendix \ref{appendixA}]. We depart from the associated Hartree-Fock uncorrelated reference state $|\psi_{HF}\rangle$ in \textit{STO-3G} basis [Tab.\ref{tab:my_label}, Appendix \ref{appendixA}], computed in JW encoding [Fig.\ref{fig:Benzene_SD}a]. In this scenario, the $\pi$ system is represented within an active space of 6 electrons, occupying the lowest energy MOs $\pi_1$,$\pi_2$ and $\pi_3$ in the HF state. Our initial objective is to identify the most relevant electronic excitations, a task that can be accomplished through an initial simulation using reduced resources. This can be achieved through various approaches, including classical computational methods or the computation of the commutator of the corresponding operators with the Hamiltonian. A notable example of the latter is used while building the ansatz in the Adaptive Derivative-Assembled Pseudo-Trotter (ADAPT) VQE \cite{Grimsley_Economou_Barnes_Mayhall_2019}. Additionally, coupled cluster methods can be employed, which are versatile and can be implemented using both classical or quantum computing techniques. In this work, we proceed with the unitary coupled cluster of singles and doubles (UCCSD), consisting in an optimization of the following parameterized state:
\begin{equation*}
|\psi_{UCCSD}\rangle=e^{\hat{\mathcal{T}}_1+\hat{\mathcal{T}}_2}|\psi_{HF}\rangle
\end{equation*}
\begin{equation*}
    \hat{\mathcal{T}}_1=\sum_{ka} \theta_{ka}\bigg(a_a^\dagger a_k-a_k^\dagger a_a\bigg)
\end{equation*}
\begin{equation}
    \hat{\mathcal{T}}_2=\sum_{klab}\theta_{klab}\bigg(a_a^\dagger a_b^\dagger a_k a_l-a_k^\dagger a_l^\dagger a_a a_b\bigg)\hspace{.25cm}.
\end{equation}
Where $k,l$ indices run occupied fermionic modes and $a,b$ unoccupied ones. More specifically, we consider the paired generalized version (UpCCGSD) \cite{Lee_2018}, which only considers double excitations promoting two electrons from two filled modes of the same MO into two other modes of another virtual MO. Additionally, single excitations are only considered between modes of equal spin, being equivalent between different spin orientations of the same MOs. From now on we define respectively the parameterized angles $\theta_{ij}^s$ and $\theta_{ij}^d$ as the ones involved in the single and double excitations between MOs $\pi_i$ and $\pi_j$. Following the UpCCGSD, we can identify relevant excitations by examining the absolute values of the optimized angles $\{|\theta_{ij}^s|\}$ for single [Fig.\ref{fig:Benzene_SD}b] and $\{|\theta_{ij}^d|\}$ for double [Fig.\ref{fig:Benzene_SD}c] excitations. 
\\
\\
In [Fig.\ref{fig:Benzene_SD}c] it becomes clear that there exist two critical double-electronic excitations. The first one (marked in red square) involves the MOs $\pi_2$ and $\pi_4$, while the second one (marked in pink square) involves MOs $\pi_3$ and $\pi_5$. Although there exist mild single electron excitations in [Fig.\ref{fig:Benzene_SD}b], the large absolute values from double excitations are more significant.
\\
\\
It is important to mention that the following process is scalable and does not require any exact diagonalization. However, when considering small molecules as in the present case, we have the privilege of accessing the exact ground state. As part of our analysis, and to illustrate how our approach simplifies the correlation structure, we compute the mutual information (MI) matrix, denoted as $I$, holding as elements $I_{ij}$ the MI between qubits $i$ and $j$. This matrix serves as a quantitative indicator of the correlation between SOs $i$ and $j$ [Fig.\ref{fig:LiH_SD}d]. An immediate observation reveals the influence of the double excitation between MO $\pi_2$ and MO $\pi_4$, as predicted above, as relevant MI is present between associated qubits 2, 3, 6, and 7 (subspace marked in red squares). The relevant double excitation between MOs $\pi_3$ and $\pi_5$ is also noticed as high correlation between associated qubits 4,5,8 and 9 (subspace marked in pink squares).

\begin{algorithm}[b]
  \caption{Algorithm employed in this work for finding reduced-entanglement fermion-to-qubit mappings.}
  \begin{algorithmic}[1]
    \Require A Hartree-Fock $|\psi_{HF}\rangle$ approximation of the target state in JW encoding, and a decision protocol $\mathcal{D}$ to select the relevant excitations from UpCCGSD analysis.\\
    Compute UpCCGSD optimization over $|\psi_{HF}\rangle$. \\
    Retrieve the absolute values for the converged angles $\{|\theta^s_{i,j}|, | \theta^d_{i,j}|\}_{i,j}$. \\
    Through the decision protocol $\mathcal{D}$, select relevant excitation/s with large value/s of $\{|\theta^s_{i,j}|, | \theta^d_{i,j}|\}_{i,j}$, involving modes with relevant electronic occupation. \\
    For each selected excitation, encode the associated fermionic modes in a $x$-branch of a TT mapping $\mathcal{T}$, replicating Eqs.(\ref{DoubleExcitation_P}, \ref{SingleExcitation_P}).\\
    Complete $\mathcal{T}$ by encoding the remaining fermionic modes occupation locally in the qubits. This is achieved by stemming a $z$-branch from the root node. \\
    \Return A tailored mapping $\mathcal{T}$ having the target state representation with reduced entanglement requirements.
  \end{algorithmic}
  \label{TailoredMappings}
\end{algorithm}

To streamline the state preparation process, the extent of long-range entanglement among the qubits can be minimized. This can be done by different means, such as minimizing the long-range MI between subparties \cite{barcza2011quantum}, or through the exchange matrix \cite{Olivares-Amaya}. Following the orbital reordering DMRG approach in the former case, we consider the cost function on our MI matrix
\begin{equation}
    C(I)=\sum_{i,j>i}I_{ij}|i-j|^2 \hspace{.25cm},
    \label{CostFunction}
\end{equation}
which relates to the Fiedler vector of the MI graph. In essence, the cost function tends to decrease when qubits with substantial MI are positioned closer to each other in a linear order. Considering the set of permutations of $n$ elements $\{P\}\in\mathcal{S}_n$, the permutation $P_{opt}$ is defined as the one that leads to a MI matrix minimizing Eq.(\ref{CostFunction}),
\begin{equation}
P_{opt}=\argmin_{P\in\mathcal{S}_m}\hspace{.1cm}C(PIP^{-1})\hspace{.25cm}.
\end{equation}
As discussed in section \ref{section2}, it is important to note that permuting physically the qubits does not alter the ground state itself but does affect the geometric constraints between subparties, which becomes crucial for upcoming optimization methods. Although finding the exact $P_{opt}$ is hard, it is sufficient a good approximation $P_s$ reducing the cost function from Eq.(\ref{CostFunction}) as much as possible. The search for $P_s$ has demonstrated remarkable benefits in improving  VQE simulations, especially on hardware with a linear restricted entangling connectivity \cite{tkachenko2021correlation}. Moreover, DMRG optimization over an MPS has also proven enhanced performance \cite{barcza2011quantum}. For the present case, we set $P_s$ as the best solution found along a genetic algorithm using DEAP \cite{DEAP_JMLR2012}. The reordered MI matrix associated to $P_s$ significantly reduces the cost function (\ref{CostFunction}), displayed in [Fig.\ref{fig:Benzene_SD}e].
\\
\\
After identifying the qubits involved in the most relevant double excitations of the system [Fig.\ref{fig:Benzene_SD}c], we proceed to transform the 4-qubit entangled subspaces resulting from the double excitations described in Eq.(\ref{DoubleExcitation_JW}) into 2-qubit entangled subspaces, as described in Eq.(\ref{DoubleExcitation_P}). To do so, we introduce the ternary tree from [Fig.\ref{fig:Benzene_SD}f], in which we have taken a decision protocol $\mathcal{D}$ that selects the modes associated to the most relevant excitations present in the UpCCGSD analysis. The qubits associated to the red/pink double excitation, are included in the red/pink $x$-branch respectively. Such tailored transformation encodes occupation non-locally between the qubits involved in each $x$-branch (similar to a parity mapping), while occupation remains locally encoded in the qubits from the $z$-branch (similar to JW mapping).
\\
\\
To visualize the effects of the tailored mapping, we examine the MI matrix of the associated ground state obtained by exact diagonalization [Fig.\ref{fig:Benzene_SD}g]. In particular, we analyze the double excitation marked in red. Qubits A/B/C/D defined in the idealized scenario from Eq.(\ref{DoubleExcitation_JW}), correspond respectively to qubits 2/3/6/7 (marked in red squares). Notably, qubits B (3) and D (7) exhibit significant entanglement in accordance with Eq.(\ref{DoubleExcitation_P}), as indicated by the mutual information elements $I_{3,7}=I_{7,3}$. The same argument holds for the pink double excitation. While the transformation has simplified the entanglement structure, it is important to remark that not all qubits have been perfectly disentangled. This is due to additional single and double excitations contributing to the overall entanglement. Most importantly, the MI matrix for the new mapping can be now rearranged minimizing the cost function from Eq.(\ref{CostFunction}), yielding [Fig.\ref{fig:Benzene_SD}h]. When we compare the value of the cost function for the reordered tailored mapping ($C=8.965$) with that of the JW-reordered matrix ($C=19.632$), it becomes evident that we have achieved a significant reduction in correlation requirements.
\\
\\
This methodology has been successfully applied to other small molecular systems [Figs.\ref{fig:LiH_SD}-\ref{fig:H22_SD}, Appendix \ref{appendixA}]. For $LiH$, the method has been used to disentangle the 0-4 double excitation. For $H_2$ in $6-31G$ basis (4 MO), the 0-1 double excitation has been simplified. In the case of the highly-correlated $H_4^{\neq}$ square system, which also featured 4 MOs and posed a 4-electron problem, we employed the same approach to simplify the double excitation involving MOs 1 and 2. In the scenario of $(H_2)_2$, we encountered two different double excitations, each involving distinct MOs. To tackle this situation, the decision protocol $\mathcal{D}$ was to select both excitations, which introduced a TT encoding scheme with a unique parity $x$-branch encoding both excitations at the same time.
\\
\\
Additionally, an stretching analysis of the $H_4^{\neq}$ Paldus system has been conducted [Fig.\ref{fig:H4_study}, Appendix \ref{appendixA}]. The MI matrix has been reconstructed for the exactly diagonalized ground state across various configurations, ranging from rectangular to square ones. The stretching parameter 
$d$ captures this variation. One side of $H_4^{\neq}$ is fixed at $1.1202$ Å, while the second side is determined by multiplying this fixed distance by the stretching parameter $d$. Across all cases explored, the tailored map yields a reduced value of the cost function, both with unordered and reordered qubits. For the square configuration ($d=1$), a single excitation from the UpCCGSD suggests the encoding of a parity branch spanning spin-orbitals 2, 3, 4, and 5. Conversely, in other configurations, three relevant double excitations emerge, reflecting the rich entanglement patterns while using canonical orbitals out of equilibrium. In this cases, we choose to encode the two double excitations from MOs 0 and 2, and from MOs 1 and 3, within distinct parity branches. Overall, our observations indicate that a suitable mapping may vary across different configurations, underscoring the importance of executing UpCCGSD for each configuration individually.
\\
\\
Using canonical orbitals instead of localized atomic orbitals disrupts the connection between resonant structures and the tree structure of the mapping. This occurs because the tree structure is linked to the correlation between canonical orbitals, whereas resonant structures are associated with the correlation between localized orbitals centered on atomic nuclei. This is particularly evident in benzene: Kekulé resonant structures relate to localized electrons on the atoms, while canonical spin-orbitals delocalize electrons across the entire molecule. A similar phenomenon is observed in $H_4^{\neq}$, where the resonant structures of horizontal and vertical bonds do not correspond to the correlations of delocalized canonical orbitals.
\\
\\
It is important to emphasize that the use of MI as a post-analysis tool was introduced in previous works. Although it is a powerful tool, its use is not strictly necessary to observe the entanglement reduction resulting from our mappings. This is evident in any of the studied cases. The MI cost function value of the non-reordered ground state for the tailored mapping (for instance [Fig.\ref{fig:Benzene_SD}g]), is smaller than that for the non-reordered JW one [Fig.\ref{fig:Benzene_SD}d].

\section{Enhancing VQE Performance}


\begin{figure}[b]                       
\centering
\includegraphics[width=1
\columnwidth, angle=0]{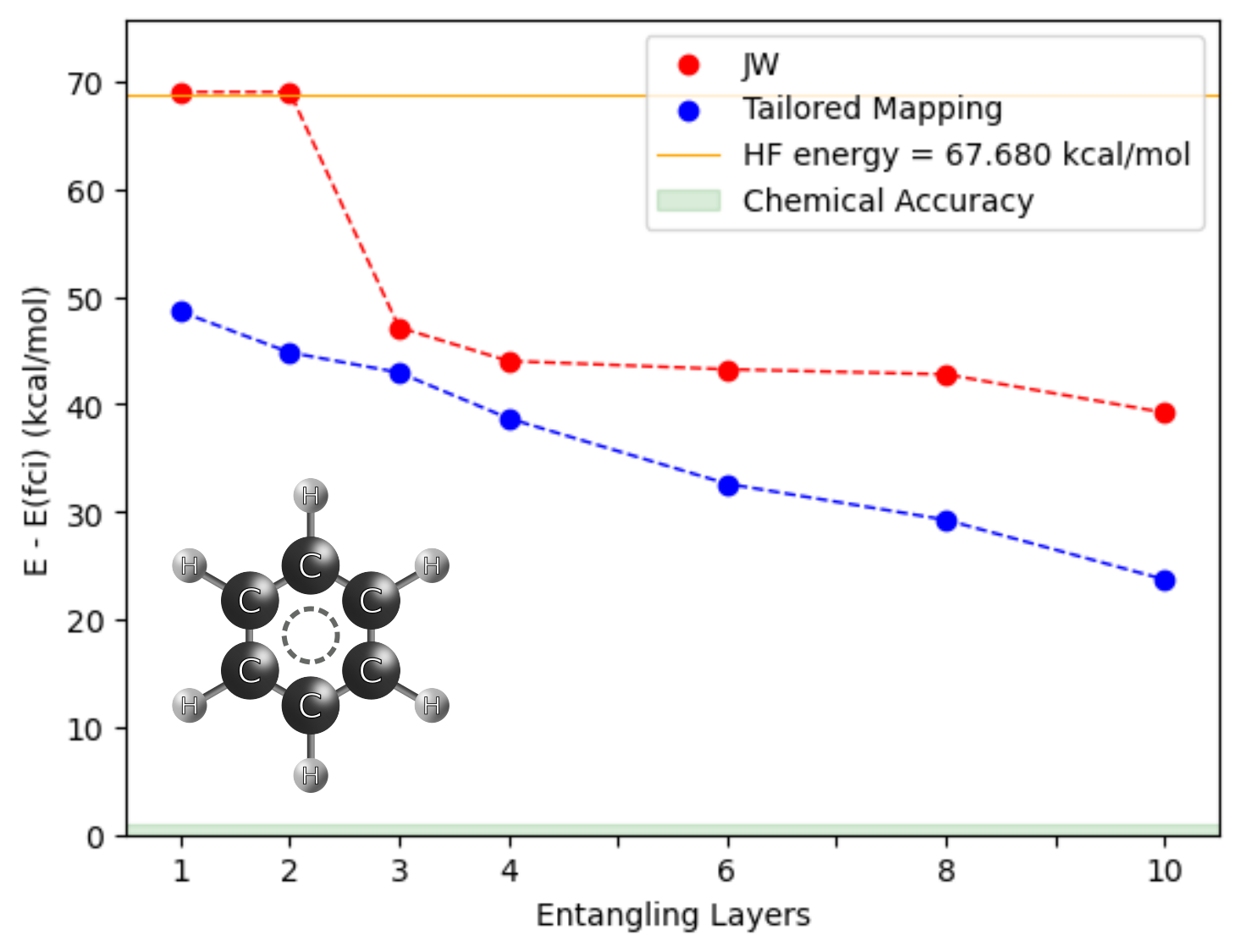}
\caption{$E_{VQE}-E_{FCI}$ as function of the RY HEA entangling layers for the $\pi$ molecular orbitals subsystem of benzene. Comparison of the JW (red) and tailored mapping (blue).}
\label{fig:Benzene_VQE}
\end{figure}

\begin{figure}[]                       
\centering
\includegraphics[width=1\columnwidth, angle=0]{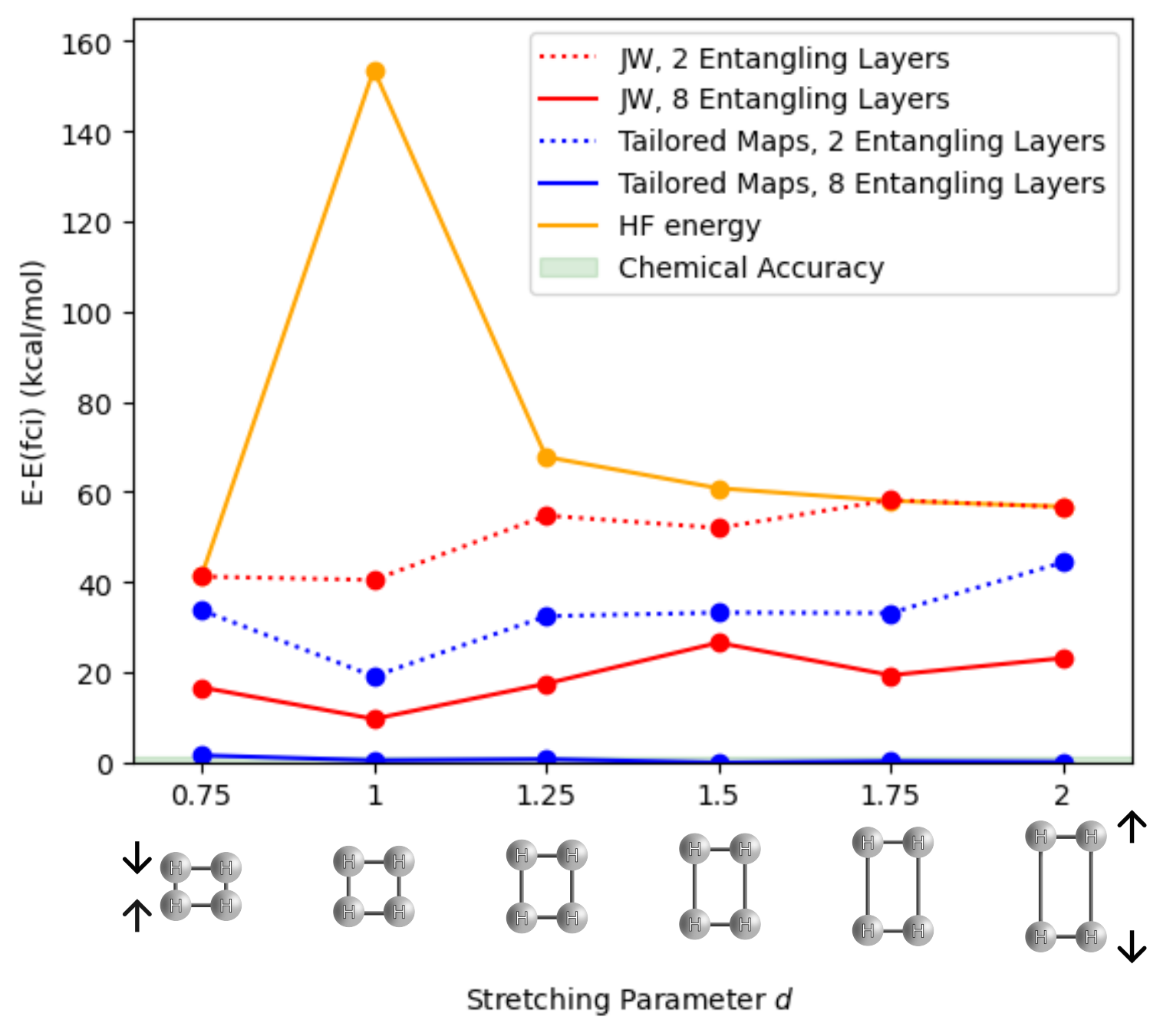}
\caption{Difference between $E_{VQE}$ and $E_{FCI}$ as a function of the stretching parameter $d$ in the stretching of $H_4^{\neq}$ Paldus system. The comparison is made between VQE results for JW (red) and tailored mappings (blue). Results are displayed for the VQE RY-HEA with 2 entangling layers (dotted lines) and 8 entangling layers (solid lines). The tailored mappings used for each stretching parameter $d$ are depicted in [Fig.\ref{fig:H4_study}, Appendix \ref{appendixA}].}
\label{fig:Paldus_VQE}
\end{figure}

On the considered molecular Hamiltonians, VQE performance has been evaluated for both JW and tailored TT mappings. VQE simulations have been implemented in \textsc{tequila} using the COBYLA optimizer \cite{powell2007view} and Qulacs simulator \cite{suzuki2021qulacs}. We compare the lowest VQE energy ($E_{VQE}$) obtained after 10 VQE executions with the FCI energy ($E_{FCI}$), with respect to the number of entangling layers employed in a HEA consisting of RY gates. The RY-HEA ansatz employed consists of alternating layers of parameterized RY rotations and entangling CNOT gates [Fig.\ref{fig:RY_HAE}, Appendix \ref{appendixA}]. 
\\
\\
In [Fig.\ref{fig:Benzene_VQE}], we display the VQE simulations for the $\pi$ system of benzene ~\cite{PhysRevA.107.012416}. The tailored mapping derived in the previous section, proves enhanced performance with respect to JW. With the JW approach, 10 entangling layers on the RY-HEA achieve a precision within 40 kcal/mol from the exact solution. In contrast, the tailored mapping achieves the same level of precision with only 4 entangling layers and offers even greater accuracy when expanded to 10 layers.
\\
\\
VQE simulations for small molecular systems are shown in [Fig.\ref{fig:VQE_exactdiag}, Appendix \ref{appendixA}], including $LiH$, $H_2$, $(H_2)_2$ and square $H_4^{\neq}$. In the figures, the benchmarked data for the JW mapping extracted from \cite{tkachenko2021correlation} is included. We conducted a replication of the method using our own hyper-parameters and number of runs, both for the JW and tailored mappings derived in the previous section. Overall, the proposed mappings demonstrate superior performance across all studied systems, leading to a faster attainment of chemical accuracy at 1 kcal/mol for the small molecules. For $LiH$, our tailored mapping achieves chemical accuracy with just 4 layers, in contrast to the 8 layers required by the JW mapping. For both $H_2$ and $(H_2)_2$ cases, the selected mapping achieves accuracy with only 6 layers, whereas JW fails to reach this level of precision even with 10 layers. For the highly-correlated $H_4^{\neq}$ square system ($E_{HF}- E_{FCI}=153.640 \hspace{.1cm} kcal/mol$), chemical accuracy is attained with the tailored mapping using 8 entangling layers, while the JW mapping falls short again when using 10 entangling layers.
\\
\\
In addition, the stretching of the Paldus $H_4^{\neq}$ system has been simulated using VQE for various stretching parameters $d$ and RY-HEA entangling layers [Fig.\ref{fig:Paldus_VQE}]. The tailored mappings, which differ for each stretching, are detailed in the correlation study for the Paldus system [Fig.\ref{fig:H4_study}, Appendix \ref{appendixA}]. As we transition to non-equilibrium geometries (such as for $d=2$), the significance of static correlations becomes evident, as demonstrated by the correlation analysis. However, the method proves to be efficient along the binding curve of $H_4^{\neq}$. When using 8 entangling layers, the tailored mappings accurately simulate any stretching up to chemical precision. In contrast, the JW mapping with an equivalent number of entangling layers, fails to attain a similar level of precision.
\\
\\
The motivation for employing a RY-HEA primitive circuit design is to provide a balanced ground for comparisons between standard and taylored fermion-to-qubit mappings. If compared to targeted circuit designs in JW representation~\cite{kottmann2023molecular}, we reach competitive circuit depths (9 vs 12 for $LiH$; 17 vs 130 for $H_4^{\neq}$). Note that by design of the RY-HEA ansatz, this comes with an increased parameter count (46 vs 5 for LiH, 64 vs 10 for H$_4$) in addition to substantially longer iteration counts. Since the techniques from~\cite{kottmann2023molecular} do not strictly require a JW representation, we estimate significant improvements in combination with the fermion-to-qubit mappings developed in this work.

\section{Enhancing DMRG Performance}

\begin{figure*}[t]                       
\centering
\includegraphics[width=2\columnwidth, angle=0]{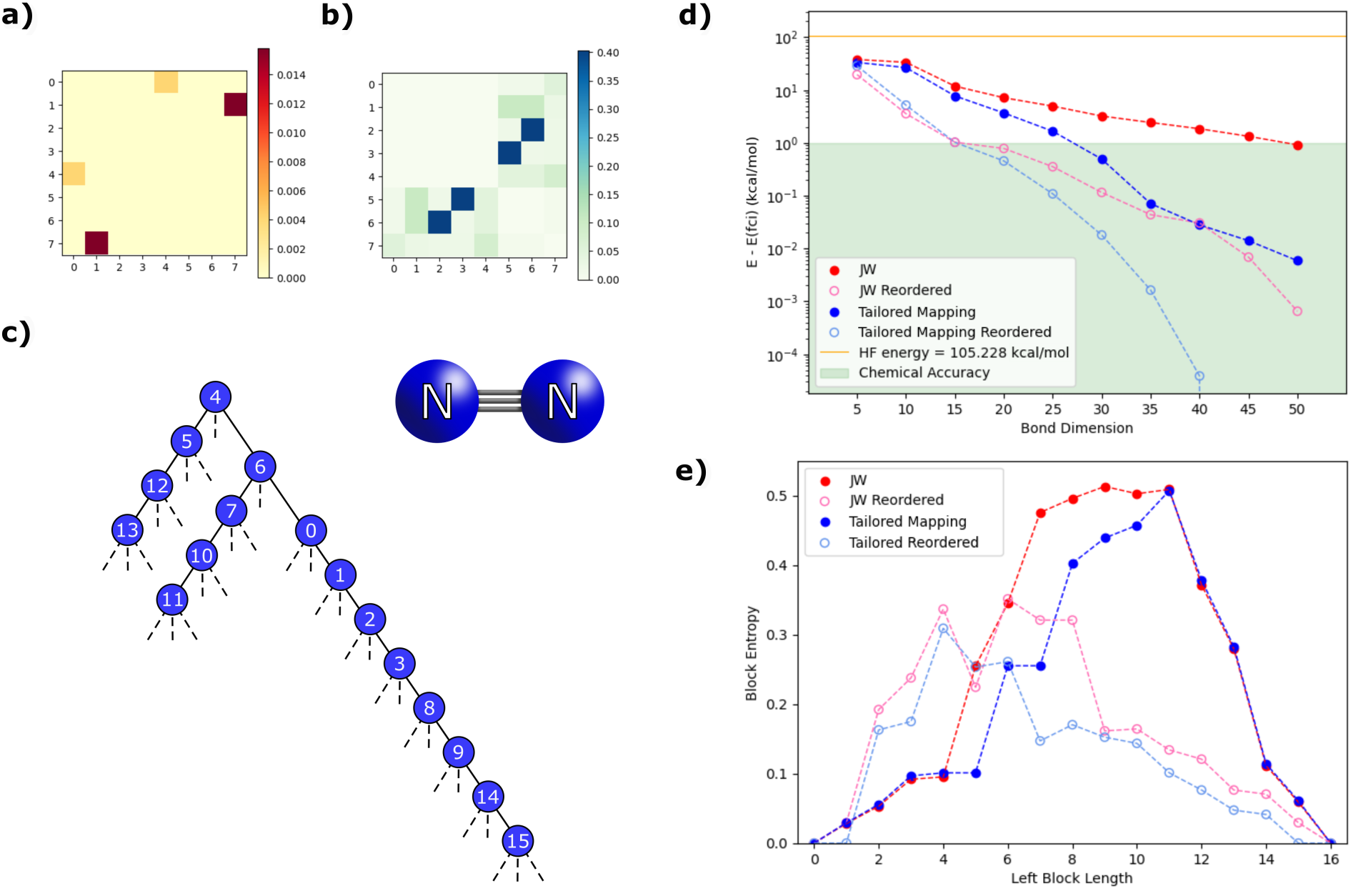}
\caption{DMRG results for the Nitrogen molecule. \textbf{a)} UpCCGSD single excitation angles between MOs for JW. \textbf{b)} UpCCGSD double excitation angles between MOs for JW. \textbf{c)} Proposed TT mapping, encoding locally in parity the spin-orbitals associated to the two double excitation subspaces $(4,5,12,13)$ and $(6,7,10,11)$. \textbf{d)} Converged $E_{DMRG}$ energy compared to $E_{FCI}$ as function of the maximum bond dimension used. The proposed TT mapping is compared to the JW mapping. The converged energy for the reordered MPS is also plotted for both mappings. \textbf{e)} Block entropies for the converged MPS after DMRG.}
\label{fig:DMRG_N2}
\end{figure*}

The methodology from the previous sections has direct applicability to tensor networks. To showcase this, we have performed a DMRG optimization over a simple low-dimensional MPS ansatz using TeNPy \cite{hauschild2018efficient}. The molecular system of $N_2$ has been initialized in \textit{STO-3G} basis with 2 core orbitals frozen, resulting in an active space of 16 SOs [Tab.\ref{tab:my_label}, Appendix \ref{appendixA}]. Again, a UpCCGSD optimization has been performed, resulting in converged angles $\{|\theta_{ij}^s|\}$ for single [Fig.\ref{fig:DMRG_N2}a] and $\{|\theta_{ij}^d|\}$ for double excitations [Fig.\ref{fig:DMRG_N2}b]. We immediately appreciate the relevant contributions of double excitations between MOs 2-6 and 3-5, and propose a TT mapping encoding such excitations in parity. In this case, we encode each double excitation in a different parity $x$-branch of the TT, as depicted in [Fig.\ref{fig:DMRG_N2}c].
\\
\\
Subsequently, a study using DMRG has been undertaken to assess the effectiveness of the proposed mapping. The energy obtained ($E_{DMRG}$) has been compared to $E_{FCI}$ for different maximum bond dimension cutoffs ($\chi_{max}$), keeping for each bond dimension the best result after 10 optimizations to avoid local minima convergence [Fig.\ref{fig:DMRG_N2}d]. For the DMRG, an exponentially-decaying mixer has been applied for 30 sweeps, allowing a maximum number of 500 sweeps for the optimization. Several scenarios have been computed to study the dependency on $\chi_{max}$. First, the best JW result is plotted in the standard spin-orbital ordering (red dots). Orbital reordering is then performed by minimizing the cost function from Eq.(\ref{CostFunction}) for the MI matrix, and DMRG is executed again in the optimized order (red circles) \cite{barcza2011quantum}. DMRG has also been performed on our tailored mapping  (blue dots), and the same reordering process has been applied to permute the MPS sites and execute DMRG again (blue circles).
\\
\\
An improvement of the DMRG performance is observed using the tailored TT mapping. For a given $\chi_{max}$, the DMRG algorithm achieves lower converged energies, leading to a more accurate simulation of the quantum state. We relate such improvements to the reduced entanglement requirements. A decrease in block entropy is observed for the tailored mapping, both in the cases of standard ordering and reordered systems. This is visualized by studying the block entropy for the best converged MPS at $\chi_{max}=50$ [Fig.\ref{fig:DMRG_N2}e]. The final MI matrices for the converged states at $\chi_{max}=50$ is displayed in [Fig.\ref{fig:N2_Converged_Dmrg}, Appendix.\ref{appendixA}].

\section{Optimality}
The presented algorithm is heuristic and based on an intuitive understanding of the role of excitations in the generation of correlations in fermion-to-qubit mappings. An enhanced performance is observed in the studied examples, however, it is not expected that our algorithm systematically provides the optimal mapping in terms of resources, say number of layers or bond dimension in the considered quantum and classical variational approaches, needed to obtain a given energy value. For $H_2$, we have studied how our construction compares to the rest of possible mappings. In particular, 
we have sampled all possible tree structures, globally optimized over the mutual information using a brute-force search of all permutations and then tested the resulting mapping with both single and double-layer VQE. For $H_2$, our tailored mapping has an optimal MI cost function of $0.611$, versus the $1.28$ of the JW map or the $1.92$ of the parity map in standard orbital ordering. Of all of the optimized MIs of all tree shapes, our tailored mapping has lower MI than 81\% of all mappings sampled [Fig.\ref{fig:optimality}, Appendix \ref{appendixA}]. This study also led to the discovery of trees with lower MI, however each of these displayed worse performance than the tailored mapping when testing with single and double-layer VQE. Although the general trend was that lower MI correlates to better few-layer VQE performance, it is shown not to be guaranteed. It should be noted that this is a heavily restricted VQE, so perhaps it is not so surprising that random mappings do not suppress the four-body correlations sufficiently in order to be expressible in the $<4$ layer VQE. 
\section{Conclusions}

In this study we have introduced a novel physically-inspired perspective in the design of fermion-to-qubit mappings. Previous efforts have predominantly focused on reducing the average Pauli weight or the number of qubits employed, which is undeniably important for quantum simulations. However, by focusing on the major sources of correlation in the physical problem, we have shown that transformations can be tailored in the qubit space to significantly simplify simulations. We remark that our formalism can be adapted both in classical and quantum computing. 
\\
\\
Our work points at several research directions that deserve further investigation. First, it is still left to consider the extension of our results to higher order fermionic excitations, on top of fermionic systems with larger spin number. Another critical aspect to investigate revolves around the correlation-mitigating capabilities of our mappings, when scaling to larger systems. Additionally, it is important to note that our method reduces circuit depth while not focusing in the Pauli weight. This suggests that our method might manifest a greater Pauli weight in comparison to alternative composable low-depth TT mappings. Subsequent efforts could be focused on optimizing this aspect while simultaneously diminishing the entanglement. In the context of noisy VQE, it emerges the question of how advantageous can this trade-off be, which can be studied in future work. 
\\
\\
Although in this work we have used archetypical simulation methods, it remains to be seen how powerful a combination of inspired mappings with other ansatzes can be. That is, in the tensor network framework, more general ansatzes differing from the linear MPS could possibly benefit widely from the reduced entanglement. In the quantum computing framework, the simple but generic RY-HEA has been used for VQE simulations. The synergy of different ansatzes with our method could be studied. These could include not only the widely recognized adaptative family, but also other approaches informed by chemical valence-bond arguments \cite{Ghasempouri_Dueck_DeBaerdemacker_2023}. Finally, while our formalism has been applied to electronic structure systems found in quantum chemistry, we anticipate possible adaptations of the method to fermionic systems of different nature, such as those found in nuclear structure or condensed matter simulations.

\section{ACKNOWLEDGEMENTS}
We thank Guillermo García Pérez for discussions. This work is supported by the ERC AdG CERQUTE, the Government of Spain (FUNQIP, NextGenerationEU PRTRC17.I1 and Quantum in Spain, Severo Ochoa CEX2019-000910-S), Fundació Cellex, Fundació Mir-Puig, Generalitat de Catalunya (CERCA programme), the AXA Chair in Quantum Information Science, EU project PASQUANS2.
\bibliographystyle{modified-apsrev4-2}
\bibliography{lit}

\clearpage

\begin{appendix}
\section{Supplementary Information}
\label{appendixA}
\begin{algorithm}[]

  \caption{Pairing algorithm for Majorana strings from a Ternary Tree. Adapted from \cite{miller2022bonsai}.}\label{alg:Majoranas}
  
  \begin{algorithmic}[1]
    \Require A ternary Tree $\mathcal{T}=(V,E)$.
    \For{$v_i \in V$}
    \If{$v_i^x$ is $None$} \Comment{Construction of $S_i^x$}
        \State Define $l$ as the $x$-downward leg from $v_i$
        \State $S_i^x\equiv S_l$
    \Else
    \State$s=v_i^x$
     \While {$s^z$ is not None}
     \State $s = s^z$
     \EndWhile
     \State Define $l$ as the $z$-downward leg from $s$
        \State $S_i^x\equiv S_l$
    \EndIf 

    \If{$v_i^y$ is $None$} \Comment{Construction of $S_i^y$}
        \State Define $l$ as the $y$-downward leg from $v_i$
        \State $S_i^y\equiv S_l$
    \Else
    \State$u=v_i^y$
     \While {$u^z$ is not None}
     \State $u = u^z$
     \EndWhile
     \State Define $l$ as the $z$-downward leg from $u$
        \State $S_i^y\equiv S_l$
    \EndIf 
    
    \EndFor
    
    \State \textbf{return} $\{S_{i}^x,S_{i}^y\}_{i}$\Comment{Paired majoranas of all nodes $v_i$}
  \end{algorithmic}
  
  \label{AlgorithmPairing}
  
\end{algorithm}

\begin{table*}[]
    \centering
    \begin{tabular}{|c|c|c c|}
    \hline
    Label & Pauli String & Ternary Tree & \\
    \hline
    \hline
    $S_0$ & $X_0Z_1Z_6$ & \multirow{21}{*}{            \includegraphics[width=1\columnwidth, angle=0]{figures/ternarytree.png}} & \\
    $S_1$ & $Y_0Z_2Z_7$ &&\\
    $S_2$ & $X_0X_1Z_4$ &&\\
    $S_3$ & $X_0Y_1Z_5$ &&\\
    $S_4$ & $Y_0X_2$ &&\\
    $S_5$ & $Y_0Y_2$ &&\\
    $S_6$ & $Z_0X_3$ &&\\
    $S_7$ & $Z_0Y_3Z_8$ &&\\
    $S_8$ & $X_0X_1X_4$ &&\\
    $S_9$ & $X_0X_1Y_4$ &&\\
    $S_{10}$ & $X_0Y_1X_5$ &&\\
    $S_{11}$ & $X_0Y_1Y_5$ &&\\
    $S_{12}$ & $X_0Z_1X_6$ &&\\
    $S_{13}$ & $X_0Z_1Y_6$ &&\\
    $S_{14}$ & $Y_0Z_2X_7$ &&\\
    $S_{15}$ & $Y_0Z_2Y_7$ &&\\
    $S_{16}$ & $Z_0Y_3X_8$ &&\\
    $S_{17}$ & $Z_0Y_3Y_8$ &&\\
    $S_{18}$ & $Z_0Z_3X_9$ &&\\
    $S_{19}$ & $Z_0Z_3Y_9$ &&\\
    $S_{20}$ & $Z_0Z_3Z_9$ &&\\
    \hline
    \end{tabular}

    \caption{Explicit Pauli strings from the ternary tree in [Fig.\ref{fig:ExampleTT}].}
    \label{tab:PauliStrings}
\end{table*}

\clearpage

\begin{table*}
    \centering
    \begin{tabular}{ |p{2cm}|p{2cm}|p{2.5cm}|p{3cm}|p{4.5cm}|  }
 \hline
 Molecule & Basis & Frozen Orbitals & $E^{corr.}(kcal/mol)$ & Cartesian Geometry (Å)\\
 \hline 
 \hline
 $H_2$   & \textit{6-31G} & 0 & 15.4945 & H  0.0000 0.0000 -0.3650 \\
   &  & & &  H  0.0000 0.0000 0.3641 \\
   \hline

 $LiH$   & \textit{STO-3G} & 1 & 12.1869 & Li  0.0000 0.0000 0.0000 \\
   &  & & &  H 0.0000 0.0000 1.5472 \\
   \hline

 $(H_2)_2$   & \textit{STO-3G} & 0 & 25.4776 & H 0.0000 0.3674 -2.1264 \\
   &  & & &  H 0.0000 -0.3674 -2.1264 \\
   &  & & &  H 0.0000 0.0000 1.7590 \\
   &  & & &  H 0.0000 0.0000 2.4939 \\
   \hline

 $H_4^{\neq} (p)$   & \textit{STO-3G} & 0 & 153.6310 (d=1) & H 0.5601 0.5601*d 0.0000\\
   &  & & &  H -0.5601 0.5601*d 0.0000  \\
   &  & & &  H -0.5601 -0.5601*d 0.0000 \\
   &  & & &  H 0.5601 -0.5601*d 0.0000  \\
   \hline
   

 $N_2$   & \textit{STO-3G} & 2 & 105.2284 & N 0.0000 0.0000 -0.5669 \\
   &  & & &  N 0.0000 0.0000 0.5669 \\
   \hline

 $C_6H_6$   & \textit{STO-3G} & 6 & $\pi$ active space: & C 0.0000 1.4027 0.0000 \\

    &  & & 67.6796& C -1.2148 0.7014 0.0000\\
    &  & & & C -1.2148 -0.7014 0.0000 \\
    &  & & & C 0.0000 -1.4027 0.0000\\
    &  & & & C 1.2148 -0.7014 0.0000\\
    &  & & & C 1.2148 0.7014 0.0000\\
    &  & & & H 0.0000 2.4901 0.0000\\
    &  & & & H -2.1567 1.2451 0.0000\\
    &  & & & H -2.1567 -1.2451 0.0000\\
    &  & & & H 0.0000 -2.4901 0.0000\\
    &  & & & H 2.1567 -1.2451 0.0000\\
    &  & & & H 2.1567 1.2451 0.0000\\
   \hline



\end{tabular}
    
    \caption{Initialization of molecular systems, including their nuclei cartesian geometry, basis set, number of core orbitals frozen and correlation energy.}
    \label{tab:my_label}
\end{table*}

\begin{figure*}[h]                       
\centering
\includegraphics[width=.8
\columnwidth, angle=0]{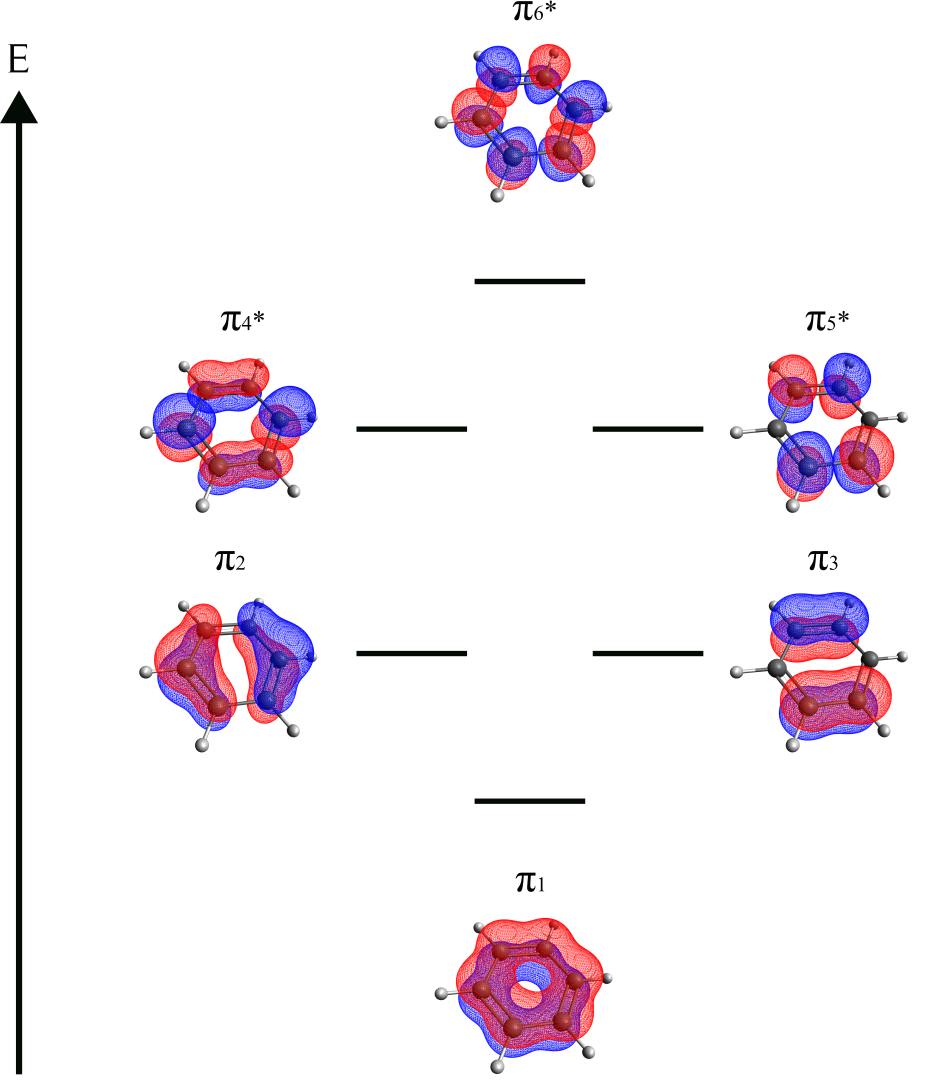}
\caption{Active space of $\pi$ molecular orbitals of benzene, expressed in the basis of canonical orbitals.}
\label{fig:Benzene_Orbitals}
\end{figure*}

\begin{figure*}                     
\centering
\includegraphics[width=2\columnwidth, angle=0]{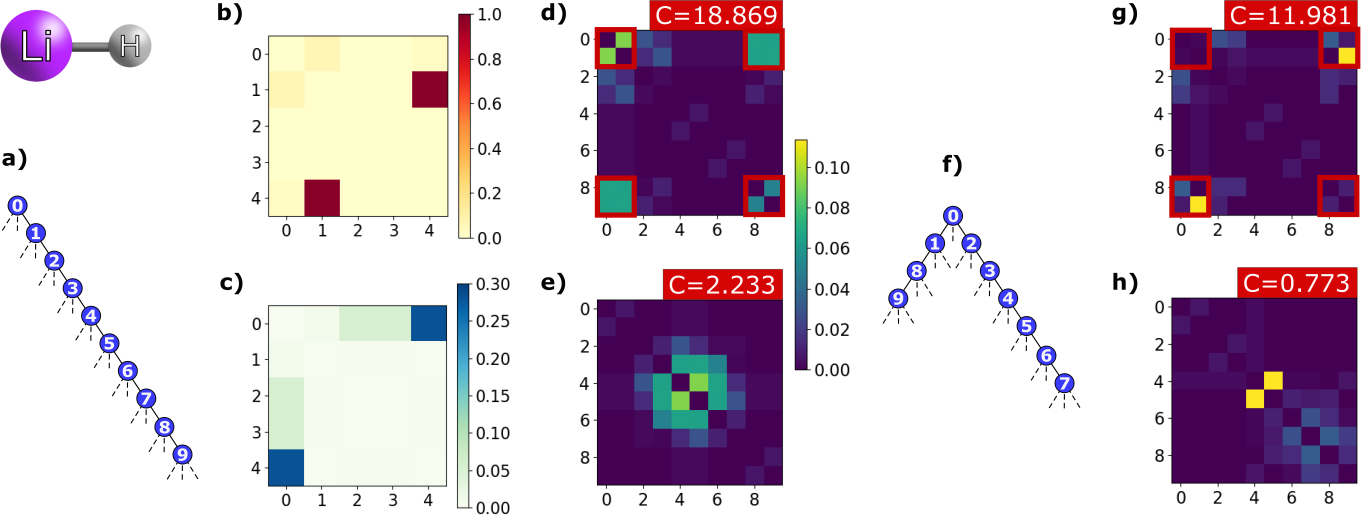}
\caption{Simplification of $LiH$ entanglement spectrum on a 5 MO active space with 2 electrons. Analogous to the benzene $\pi$ system in [Fig.\ref{fig:Benzene_SD}]. An analysis in the JW encoding is performed in first instance. The gathered information is then used to construct a tailored mapping. \textbf{a)} TT corresponding to JW encoding \textbf{b)} Absolute values $|\theta_{ij}^s|$ after UpCCGSD, corresponding to single excitations between MOs $i$ and $j$. \textbf{c)} Absolute values $|\theta_{ij}^d|$ after UpCCGSD, corresponding to double excitations between MOs $i$ and $j$. \textbf{d)} Mutual information matrix between SOs for JW ground state. \textbf{e)} Mutual information matrix reordered minimizing Eq.(\ref{CostFunction}). \textbf{f)} Constructed mapping encoding the most relevant double excitation subspace locally in parity. \textbf{g)} Ground state mutual information matrix for the adapted mapping that reduces entanglement on the double excitation between SOs 0,1,2 and 3. \textbf{h)} Ground state reordered mutual information matrix for the adapted mapping. Red labels indicate the cost function value of the corresponding mutual information matrices, and red squares indicate the 4 qubits directly involved in the most relevant double excitation.}
\label{fig:LiH_SD}
\end{figure*}

\begin{figure*}                      
\centering
\includegraphics[width=2\columnwidth, angle=0]{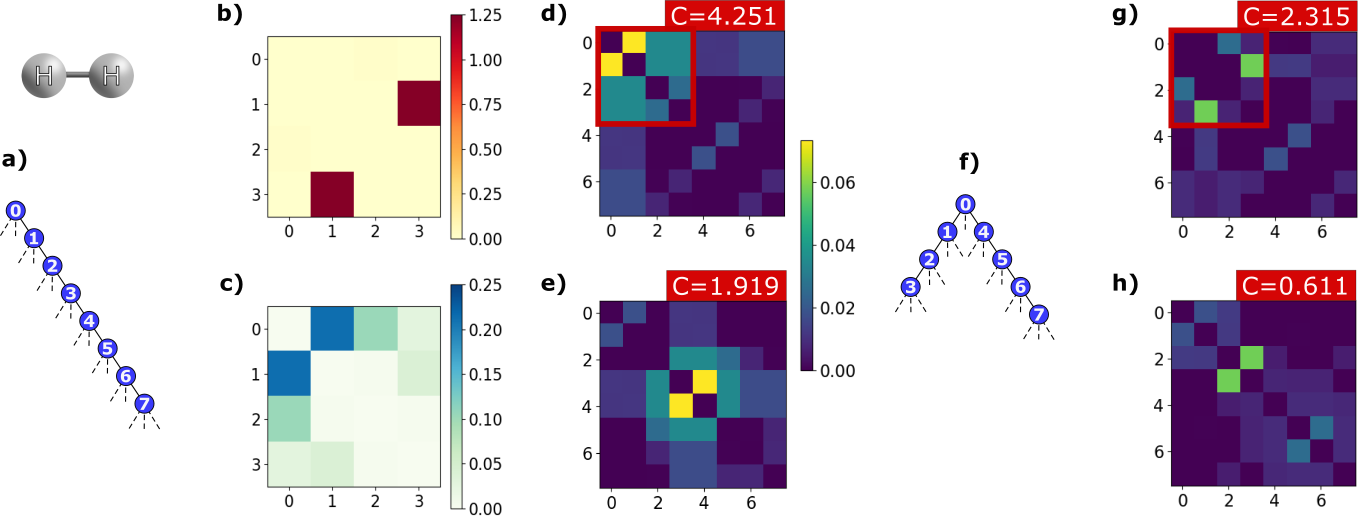}
\caption{Entanglement reduction process for $H_2$. Analogous to the benzene $\pi$ system in [Fig.\ref{fig:Benzene_SD}]. \textbf{a)} JW mapping. \textbf{b)} Absolute values $|\theta_{ij}^s|$ after UpCCGSD. \textbf{c)} Absolute values $|\theta_{ij}^d|$ after UpCCGSD. \textbf{d)} MI matrix between SOs for JW ground state. SOs involved in the most relevant double excitation are marked in the red square. \textbf{e)} Reordered JW MI matrix minimizing Eq.(\ref{CostFunction}). Cost is indicated in red. \textbf{f)} Tailored mapping proposed from the excitation analysis. \textbf{g)} Ground state MI matrix for the tailored mapping that simplifies entanglement on the qubits involved in the relevant double excitation (marked in the red square). \textbf{h)} Ground state reordered MI matrix for the tailored mapping. The cost function is indicated in the red labels.}
\label{fig:H2_SD}
\end{figure*}

\begin{figure*}                     
\centering
\includegraphics[width=2\columnwidth, angle=0]{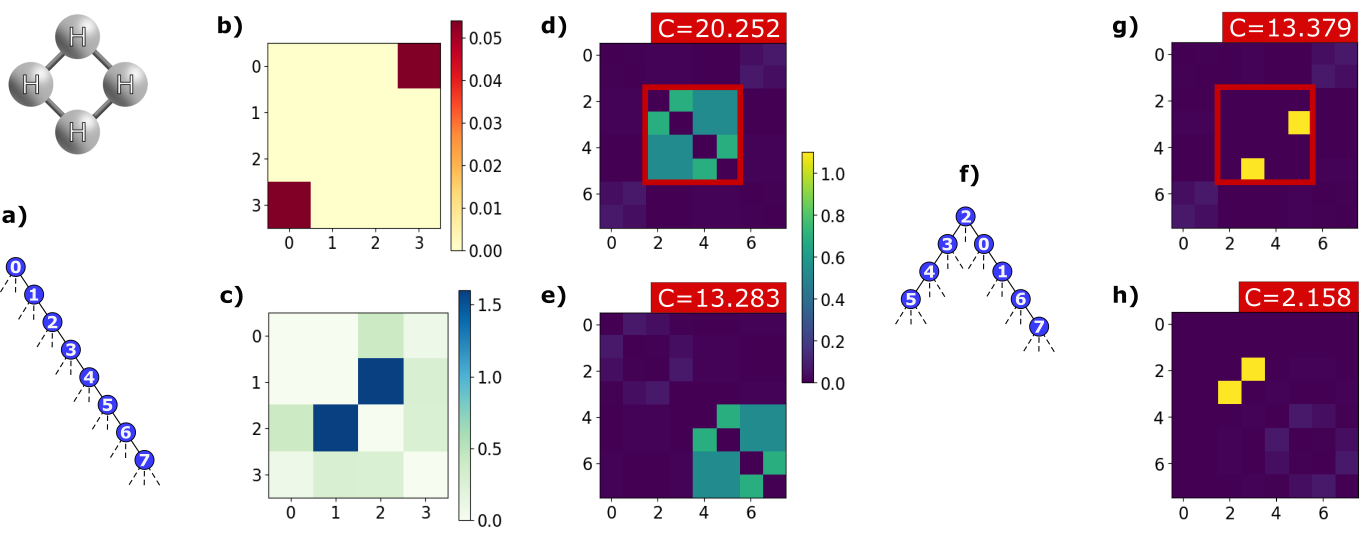}
\caption{Entanglement reduction process for square $H_4^{\neq}$. Analogous to the benzene $\pi$ system in [Fig.\ref{fig:Benzene_SD}]. \textbf{a)} JW mapping. \textbf{b)} Absolute values $|\theta_{ij}^s|$ after UpCCGSD. \textbf{c)} Absolute values $|\theta_{ij}^d|$ after UpCCGSD. \textbf{d)} MI matrix between SOs for JW ground state. SOs involved in the most relevant double excitation are marked in the red square. \textbf{e)} Reordered JW MI matrix minimizing Eq.(\ref{CostFunction}). Cost is indicated in red. \textbf{f)} Tailored mapping proposed from the excitation analysis. \textbf{g)} Ground state MI matrix for the tailored mapping that simplifies entanglement on the qubits involved in the relevant double excitation (marked in the red square). \textbf{h)} Ground state reordered MI matrix for the tailored mapping. The cost function is indicated in the red labels.}
\label{fig:H4_SD}
\end{figure*}

\begin{figure*}                       
\centering
\includegraphics[width=2\columnwidth, angle=0]{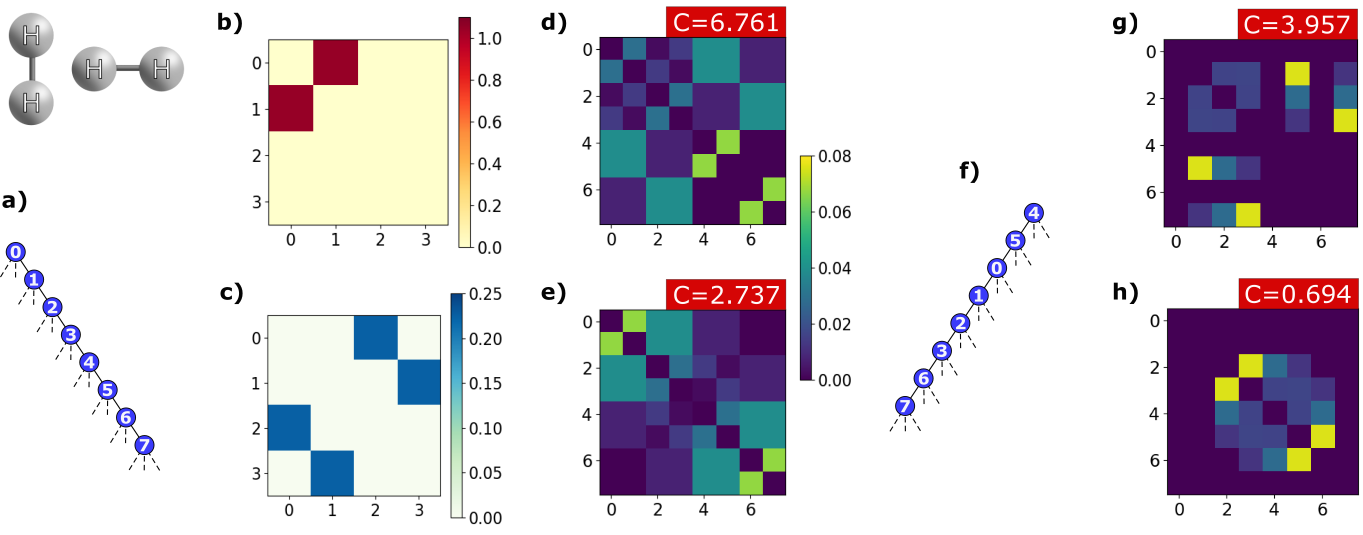}
\caption{Entanglement reduction process for $(H_2)_2$. Analogous to the benzene $\pi$ system in [Fig.\ref{fig:Benzene_SD}]. \textbf{a)} JW mapping. \textbf{b)} Absolute values $|\theta_{ij}^s|$ after UpCCGSD. \textbf{c)} Absolute values $|\theta_{ij}^d|$ after UpCCGSD. \textbf{d)} MI matrix between SOs for JW ground state. SOs involved in the most relevant double excitation are marked in the red square. \textbf{e)} Reordered JW MI matrix minimizing Eq.(\ref{CostFunction}). Cost is indicated in red. \textbf{f)} Tailored mapping proposed from the excitation analysis. Note that the two double excitations involving SOs (0,1,4,5) and (2,3,6,7) are included contiguously in the resulting parity order, on top of the (0,1,2,3) SOs involved in the single excitation. \textbf{g)} Ground state MI matrix for the tailored mapping that simplifies entanglement on the qubits involved in the relevant double excitation (marked in the red square). \textbf{h)} Ground state reordered MI matrix for the tailored mapping. The cost function is indicated in the red labels.}
\label{fig:H22_SD}
\end{figure*}


\begin{figure*}                       
\centering
\includegraphics[width=2\columnwidth, angle=0]{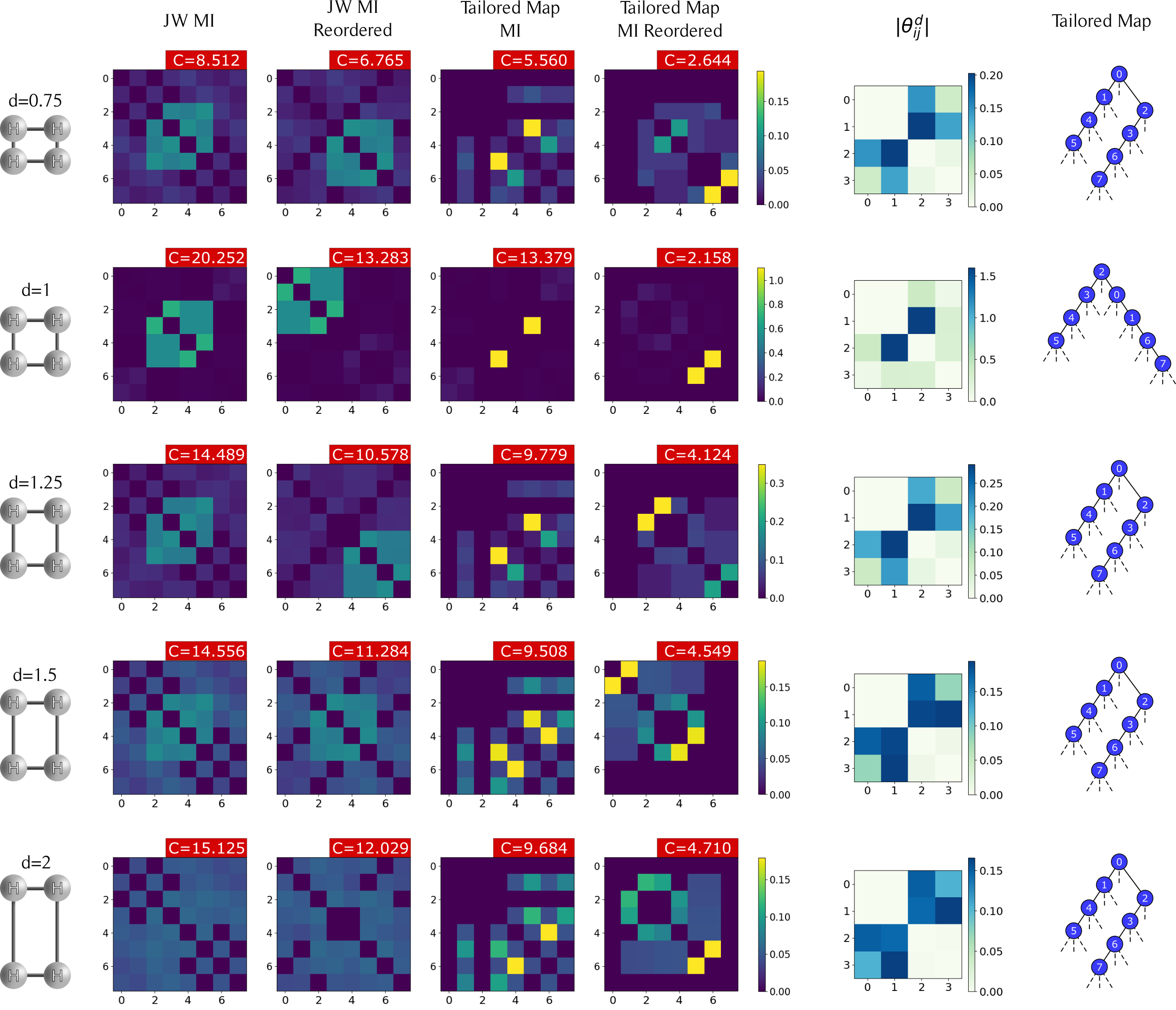}
\caption{Ground state mutual information of the Paldus rectangle-to-square transition of $H_4^{\neq}$. Each row represents a distinct relationship $d$ between side lengths, with a fixed horizontal length of $1.1203$ Å. The first two columns depict the MI matrices of the JW map (unordered and reordered, respectively). The third and the fourth columns represent the MI matrices from the tailored map (unordered and reordered, respectively). The tailored map is built from the double excitation analysis for each case in the fifth column, and is explicitly shown at the last column. The cost function for each MI matrix is indicated in the red labels. Single excitation angles are omitted from display as they hold reduced significance within the analyzed system.}
\label{fig:H4_study}
\end{figure*}

\begin{figure*}[t]                       
\centering
\includegraphics[width=2\columnwidth, angle=0]{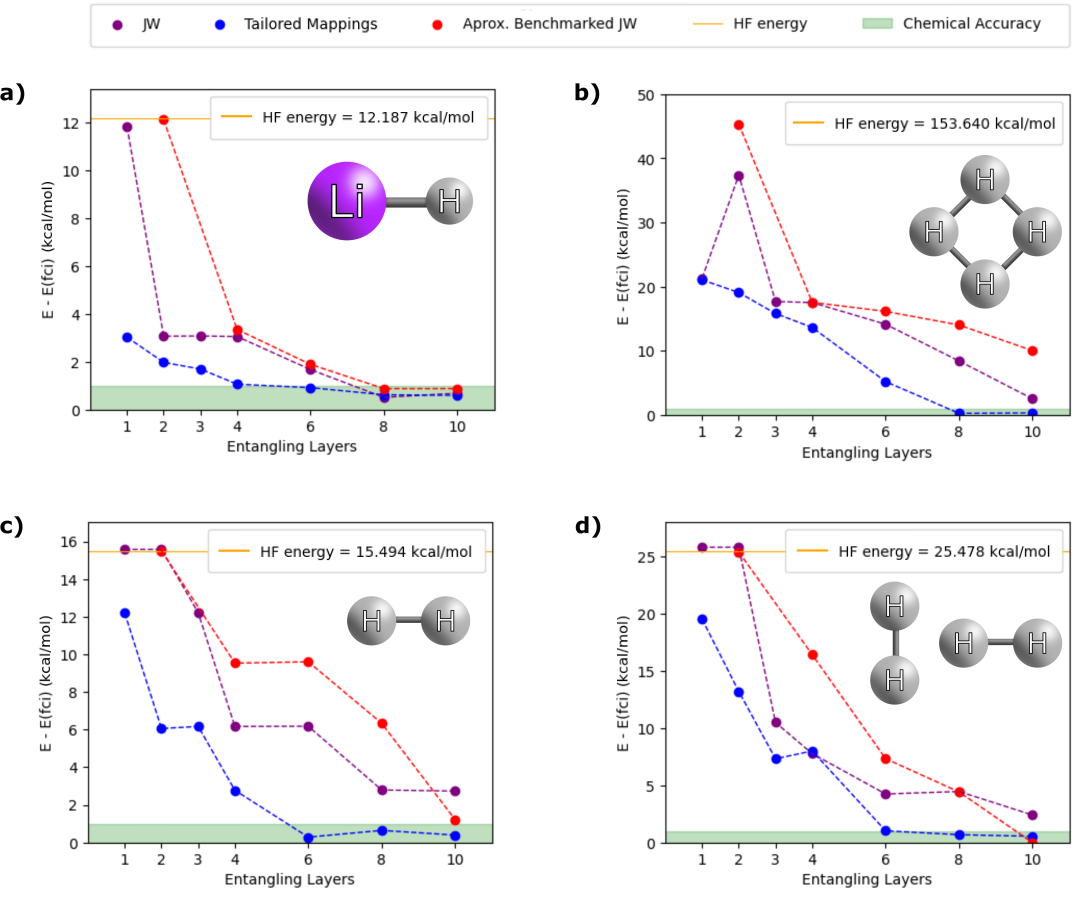}
\caption{$E_{VQE}-E_{FCI}$ as function of the RY HEA entangling layers for the considered molecular systems of \textbf{a)}$LiH$ \textbf{b)}$H_4^{\neq}$ \textbf{c)}$H_2$ \textbf{d)}$(H_2)_2$. Comparison of the benchmarked JW reordered (red) in \cite{tkachenko2021correlation}, JW reordered executed with our optimizers, backends and hyperparameters for fair comparison (purple) and tailored mapping reordered (blue).}

\label{fig:VQE_exactdiag}
\end{figure*}

\begin{figure*}                       
\centering
\includegraphics[width=1.2\columnwidth, angle=0]{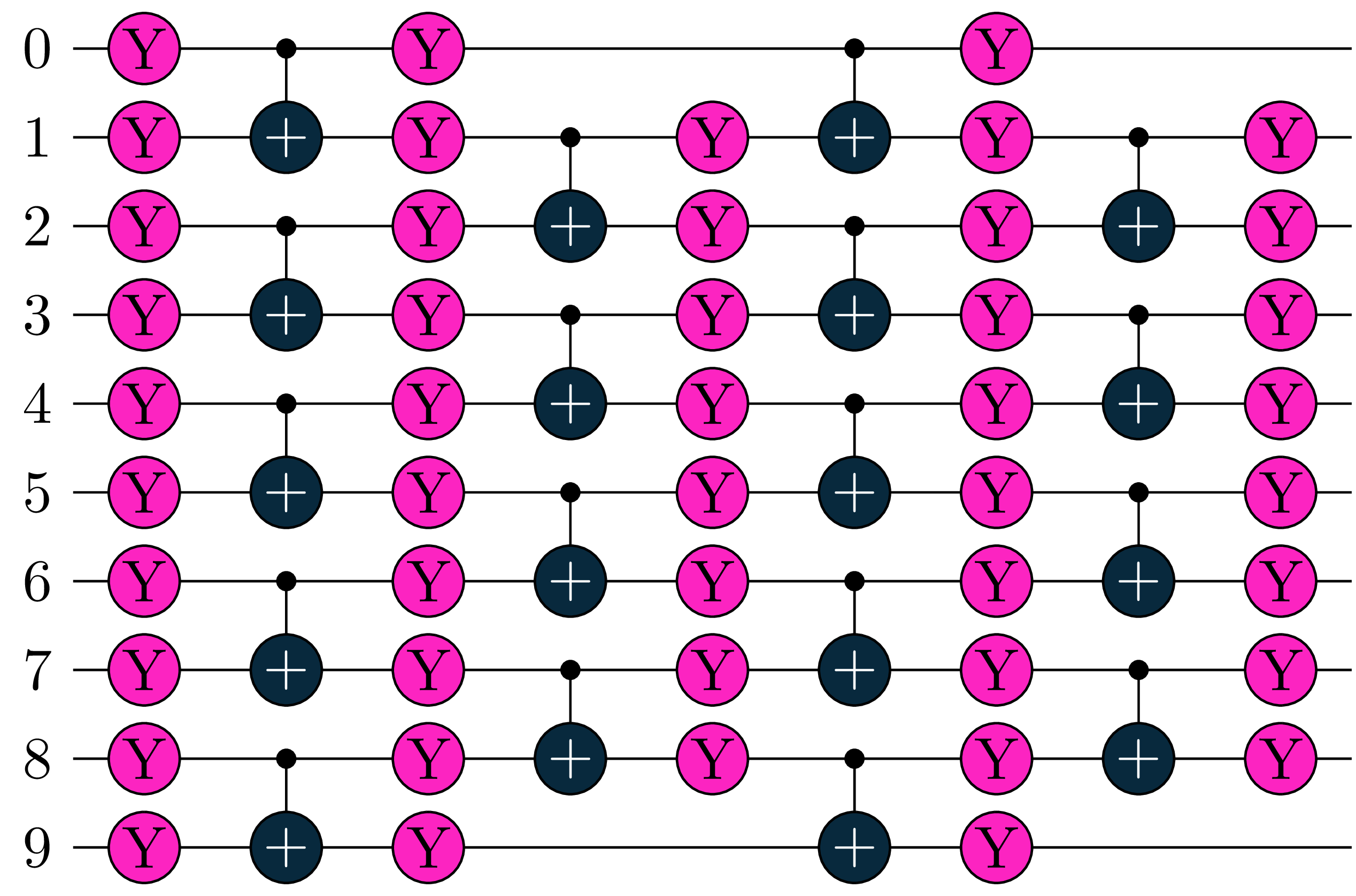}
\caption{Example of parameterized RY-HEA ansatz of 4 entangling layers,  assembled from parameterized $R_y$ rotations and static \textsc{CNOT}s, used for the VQE optimization. }
\label{fig:RY_HAE}
\end{figure*}

\begin{figure*}                       
\centering
\includegraphics[width=1.5\columnwidth, angle=0]{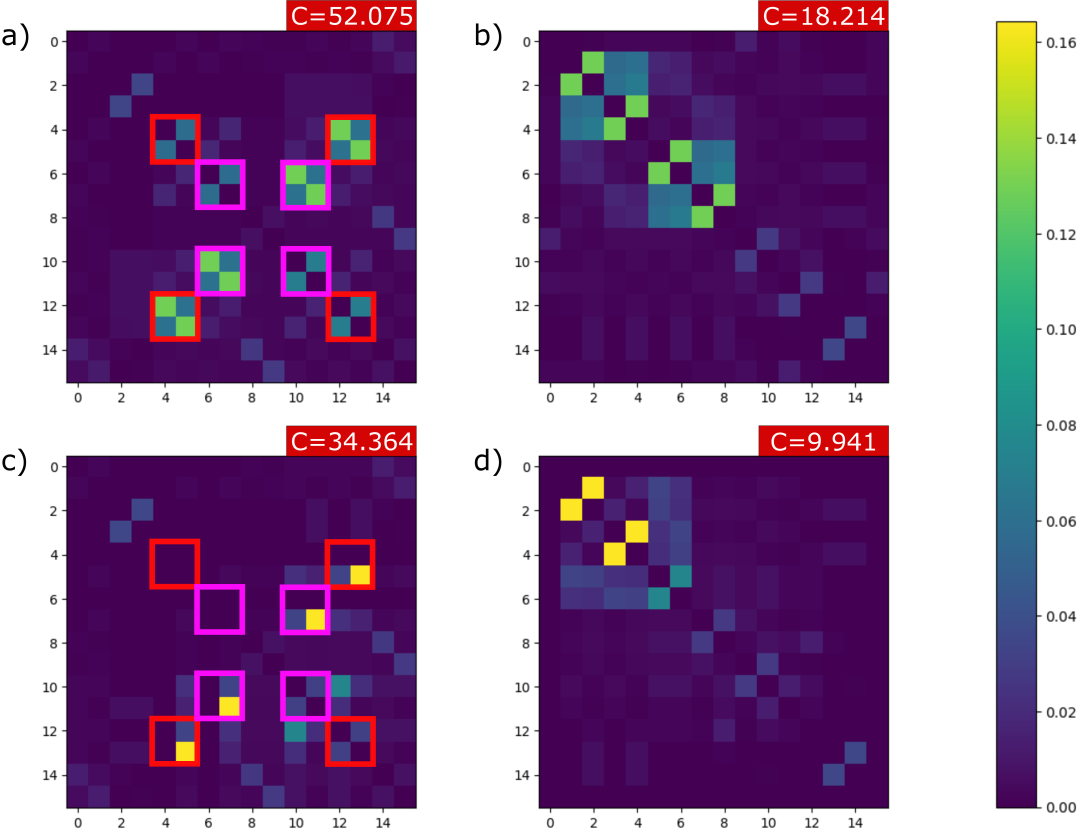}
\caption{N2 MI matrices for the converged MPS after DMRG for maximum bond dimension 50 in [Fig.\ref{fig:DMRG_N2}]. The cost function values are indicated in the red labels. The qubits involved in the two relevant double excitations are marked in red and pink squares. \textbf{a)} JW encoding.  \textbf{b)} JW reordered.  \textbf{c)} Constructed mapping after excitation analysis in [Fig.\ref{fig:DMRG_N2}c].  \textbf{d)} The same constructed mapping reordered. }
\label{fig:N2_Converged_Dmrg}
\end{figure*}

\begin{figure*}                       
\centering
\includegraphics[width=1.2\columnwidth, angle=0]{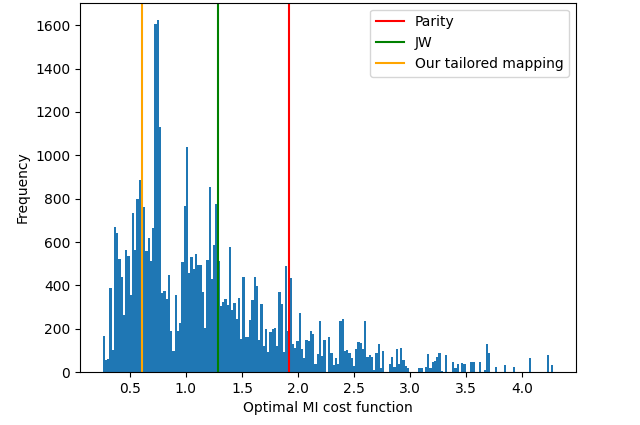}
\caption{Distribution of MI cost function for the sampling of mappings in the optimality section for the $H_2$ molecule.}
\label{fig:optimality}
\end{figure*}


\end{appendix}

\end{document}